\documentclass[aps,preprint,notitlepage, superscriptaddress]{revtex4-2}

\usepackage{amsmath,amssymb,amsfonts}
\usepackage{graphicx}
\usepackage{booktabs}
\usepackage{hyperref}
\usepackage{xcolor}
\usepackage{enumitem}
\usepackage{listings}
\usepackage{tikz}
\usepackage{braket}
\usepackage{float}
\usepackage{ulem}

\definecolor{codeblue}{RGB}{0,102,204}

\hypersetup{
    colorlinks=true,
    linkcolor=codeblue,
    citecolor=codeblue,
    urlcolor=codeblue,
}

\begin{document}

\title{Real-Time Quantum Error Correction System Stack: Architecture, Algorithms, and Engineering Practice}

\author{Yaojian Chen}
\affiliation{FieldQuantum Research Institute, Beijing 100176, China}
\affiliation{FieldQuantum Frontier Research Institute, Beijing 100190, China}
\author{Chun-Yang Luan}
\affiliation{China Mobile Research Institute, Beijing 100032, China}
\author{Peilin Zheng}
\affiliation{China Mobile Research Institute, Beijing 100032, China}
\author{Xianghong Zeng}
\affiliation{Xoherence Co., Ltd}
\author{Jia-Yi Hou}
\affiliation{CAS Cold Atom Technology (Wuhan) Co., Ltd., Wuhan 430073, China}
\author{Zhuo Fu}
\email{fuzhuo@cascoldatom.com}
\affiliation{CAS Cold Atom Technology (Wuhan) Co., Ltd., Wuhan 430073, China}
\author{Yirong Jin}
\email{jinyr@xoherence.com}
\affiliation{Xoherence Co., Ltd}
\author{Fei Wang}
\email{wangfei@chinamobile.com}
\affiliation{China Mobile Research Institute, Beijing 100032, China}
\author{Guangwen Yang}
\affiliation{Beijing Taichu Yuanxin Integrated Circuit Co., Ltd., Beijing 100176, China}
\author{Dingshun Lv}
\email{lvdingshun@fieldquanta.com}
\affiliation{FieldQuantum Research Institute, Beijing 100176, China}
\affiliation{FieldQuantum Frontier Research Institute, Beijing 100190, China}

\begin{abstract}
Quantum error correction (QEC) is transitioning from physical feasibility demonstrations to systems engineering challenges.
Google has achieved below-threshold performance on distance-5/7 surface codes, while Riverlane and Rigetti have demonstrated hardware-integrated low-latency feedback loops.
These milestones indicate that the core challenge of real-time decoding has shifted from algorithmic capability to system-level engineering.
However, a substantial engineering gap remains between laboratory demonstrations and scalable fault-tolerant quantum computing (FTQC).
This white paper addresses three questions:
(1) Where are the real bottlenecks in real-time QEC: beyond average decoder speed, the constraints lie in QEC round time, tail latency, and end-to-end data path coordination;
(2) How mature are mainstream decoder algorithms: we benchmark the major decoders for both surface codes and quantum low-density parity-check (qLDPC) codes, evaluating their real-time readiness;
(3) What system stack do we propose: a six-layer reference architecture from syndrome acquisition to logical operations, with interface definitions and latency budget models.
Our results quantify the gap between current decoder performance and real-time requirements, and identify the architectural choices needed to close it.
\end{abstract}

\maketitle
\newpage
\tableofcontents

\section{Introduction}
\label{sec:introduction}

Quantum computing offers potential speedups for selected problems in cryptography, quantum chemistry, materials science, and optimization, yet current quantum hardware remains fundamentally limited by noise~\cite{preskill2018nisq}.
Quantum error correction (QEC) provides the theoretical foundation for fault-tolerant quantum computing (FTQC) by encoding logical qubits into redundant physical qubits and continuously detecting and correcting errors.
After decades of theoretical development, QEC has now crossed early experimental milestones toward fault tolerance.
In 2025, Google demonstrated exponential suppression of logical error rates with increasing code distance on the Willow processor using rotated surface codes of distances $d=3, 5, 7$~\cite{google2025threshold}.
Riverlane and Rigetti achieved hardware-integrated real-time decoding with sub-microsecond latency in a simplified surface code with $8$ qubits~\cite{barber2025realtime}.
Quantinuum demonstrated the first fault-tolerant logical teleportation and lattice surgery via real-time active QEC using the $[[7,1,3]]$ color code (Steane code) on trapped ions~\cite{ryan2024high}, subsequently achieving a break-even point with logical error rates 800 times lower than physical baselines~\cite{microsoft2024break} and executing fault-tolerant algorithms on up to 12 logical qubits~\cite{quantinuum2026algorithms}.
QuEra operated a programmable logical processor on reconfigurable neutral atom arrays with up to 48 logical qubits using error-detected logical encoding without active correction~\cite{bluvstein2024logical}.
These progresses indicate that the open question is now how to engineer them into a reliable, scalable system.

The gap between a laboratory demonstration and a production-grade fault-tolerant quantum computer is largely an engineering gap.
A distance-7 surface code experiment running for a few dozen syndrome rounds differs profoundly from a distance-17 system that must operate continuously for millions of rounds, where the latter must manage backlog, coordinate multiple logical qubits, and meet hard latency deadlines for non-Clifford gates.
Bridging this gap demands a systematic engineering perspective that looks beyond algorithmic performance to encompass the full data path, architectural choices, and integration constraints.

In this white paper, we analyze the full data path from syndrome measurement to logical operation, benchmark the major decoder algorithms under realistic conditions, and propose a layered architecture that can evolve with both hardware and algorithmic advances.
Our goal is to identify the practical bottlenecks and architectural decisions that will determine whether fault-tolerant quantum computing scales.

\subsection{Thesis}
\label{subsec:thesis}

The following six claims form the central thesis of this paper.
Each is substantiated in subsequent chapters with technical analysis and empirical data.

\begin{enumerate}

\item \textbf{The bottleneck has shifted from algorithms to systems.}
State-of-the-art decoders meet real-time deadlines on average for small $d$, but meeting them continuously and reliably under full system constraints, including sustained throughput, bounded tail latency, and deterministic feedback, remains an open engineering challenge~\cite{terhal2015online}.

\item \textbf{Most closed loops do not involve active physical correction.}
The majority of QEC experiments operate in a Pauli frame tracking mode: corrections are accumulated classically rather than applied physically~\cite{fowler2012surface}.
However, the relevant decoded frame information must be available before any feedforward-controlled operations can be determined, such as adaptive measurements, conditional Clifford corrections, and magic-state consumption.
In other words, this resolution requires real-time classical feedback with hard latency deadlines.
These two feedback modes impose fundamentally different latency requirements on the decoder.

\item \textbf{A decoder that falls behind has compounding consequences.}
When the sustained decoder throughput cannot keep pace with syndrome generation, the consequences appear most sharply at synchronization barriers.
First, decoder-induced synchronization stalls force data qubits to idle, allowing additional physical errors to accumulate.
In this case, the accumulated idle errors effectively raise the physical error rate, potentially pushing it above the code threshold.
Second, backlog stalls logical operations.
Non-Clifford operations and other adaptive steps may have to wait until the decoder has caught up, directly inflating algorithm runtime~\cite{terhal2015online}.
These effects make sustained decode throughput a hard system requirement, not merely a performance target.

\item \textbf{Tail latency is the operationally relevant metric.}
A decoder that is fast on average but occasionally stalls (e.g., 200\,ns typical but 50\,$\mu$s worst-case) can accumulate backlog and force additional QPU idle time.
Optimizing for average throughput while ignoring tail behavior gives a misleading picture of decoder readiness.

\item \textbf{Surface codes and qLDPC codes impose very different system requirements.}
Surface codes have local stabilizers, mature decoders, and well-understood integration paths~\cite{fowler2012surface}.
Quantum low-density parity-check (qLDPC) codes (such as bivariate bicycle codes~\cite{bravyi2024high}) offer better data-qubit encoding rates but bring harder layout constraints, deeper syndrome extraction circuits, additional ancilla and measurement-resource requirements, and slower decoders in practice.
A practical system stack must support both.

\item \textbf{Architecture and computer system engineering will be the next differentiators.}
As decoder algorithms mature, competitive advantage shifts to heterogeneous compute architecture (FPGA + GPU + CPU), performance engineering at large code distances ($d \geq 17$), and end-to-end system integration~\cite{barber2025realtime, chamberland2026ising}.
These are areas where computer system expertise becomes decisive.

\end{enumerate}

\subsection{Scope and structure}
\label{subsec:scope}

This paper is organized as follows.
Chapter~\ref{sec:background} surveys the industry and technical landscape, establishing the experimental milestones that motivate our systems focus.
Chapter~\ref{sec:system_model} formalizes the system model and defines the latency, throughput, and reliability metrics we use throughout, including platform-dependent real-time constraints for superconducting, trapped-ion, and neutral-atom systems.
Chapters~\ref{sec:codes} and~\ref{sec:decoders} review the error-correcting codes and decoder algorithms relevant to real-time QEC, including hardware-code co-design considerations for different platforms.
Chapter~\ref{sec:benchmark} presents our benchmark results on surface codes and qLDPC codes using \texttt{stim} circuit-level noise simulations.
Chapter~\ref{sec:architecture} proposes a six-layer reference architecture for the real-time QEC system stack, specifies the structured decoder interface used by metadata-rich platforms, and discusses hardware acceleration, scaling strategies, and platform-specific stack implementations.
Chapter~\ref{sec:applications} traces the path from decoder performance to application-level impact and identifies the critical middleware gaps in the current ecosystem.
\section{Technical and Industry Background}
\label{sec:background}

\subsection{Quantum error correction in a nutshell}
\label{subsec:qec_basics}

A quantum error-correcting code $[[n, k, d]]$ encodes $k$ logical qubits into $n$ physical qubits with code distance $d$, meaning it can correct any error affecting up to $\lfloor (d-1)/2 \rfloor$ qubits.
The code is defined by a set of commuting Pauli operators called stabilizers.
Measuring all stabilizers projects the system onto a joint eigenstate without disturbing the encoded information, producing a classical bit string called the syndrome.
In the absence of errors the syndrome is trivial (all $+1$), while any nontrivial syndrome pattern signals the presence of one or more errors.

\paragraph{The QEC round.}
A simplified model for real-time QEC is described as a repeated cycle of five steps:
(i)~entangle ancilla qubits with data qubits via a stabilizer measurement circuit,
(ii)~measure the ancillas to extract the syndrome,
(iii)~transmit the syndrome to a classical decoder,
(iv)~infer the most likely error and return a correction,
(v)~apply or record the correction~\cite{barber2025realtime}.
One complete iteration is called a QEC round, and a distance-$d$ surface code typically executes $O(d)$ rounds to build up sufficient temporal redundancy against measurement errors.

Notably, this per-round description is a simplified model.
In practice, the decoder does not need to finish decoding each round immediately.
Instead, it is a continuous classical process that consumes a stream of syndrome data and may produce corrections with non-zero latency.
Moreover, it does not operate on each round syndrome data in isolation, but rather processes multiple rounds as a batch~\cite{google2025threshold} or uses a sliding-window approach that commits corrections incrementally~\cite{skoric2023parallel}.

During Clifford-only computation, corrections need not be physically applied.
Clifford gates (such as H, S, CNOT) map Pauli operators to Pauli operators under conjugation ($C E C^\dagger \in \mathcal{P}_n$ for any Pauli $E$), so corrections can be commuted through them and accumulated in a classical data structure called the Pauli frame, where no physical intervention is needed~\cite{fowler2012surface}.
This means the decoder may lag behind by many rounds without affecting the computation.
A hard synchronization deadline arises when subsequent operations require resolved frame information, for example, during adaptive measurement-basis selection or conditional Clifford correction associated with non-Clifford gates.
The key requirement is that the relevant decoded frame information must be available before the corresponding adaptive decision is fixed.
At the barrier, unfinished decoding directly stalls the logical operation.

\paragraph{QEC and logical operations.}
In practice, QEC runs continuously throughout the computation.
The stabilizer measurement circuit repeats every round, and logical operations are realized by modifying the stabilizer schedule, the code patch geometry, or the physical gate layer for a certain number of rounds.
How QEC interleaves with computation depends on the gate implementation method.

\textit{State preparation and encoding.}
A logical qubit is initialized by preparing the physical qubits in a product state and then projecting onto the code space using $d$ rounds of stabilizer measurements.
The decoder processes these initial rounds to verify that the state was correctly prepared.

\textit{Transversal gates.}
A transversal gate applies a physical gate independently to each qubit (or each qubit pair).
The transversal gate is executed within a single QEC round (the stabilizer circuit may be briefly modified), after which normal QEC resumes.
The decoder treats this round like any other syndrome round and no additional synchronization is needed provided the gate is a Clifford operation.

\textit{Lattice surgery.}
For surface codes, most two-qubit Clifford operations are implemented via lattice surgery~\cite{horsman2012surface}.
Logical CNOT, for example, is performed by temporarily merging two surface code patches (measuring joint stabilizers across the boundary for $d$~rounds) and then splitting them (restoring the original stabilizers for another $d$~rounds).
During both phases, the stabilizer measurement circuit continues to run, with a sequence of modified QEC rounds.
After any lattice surgery operation, $O(d)$ additional idle QEC rounds are typically performed to re-establish confidence in the error state before the next operation~\cite{fowler2012surface, litinski2019magic}.

\textit{Magic-state distillation.}
T~gates (and other non-Clifford gates) cannot be implemented transversally on most codes.
The standard approach is magic-state distillation~\cite{bravyi2005universal}:
(i)~prepare a magic state $\ket{T} = (|0\rangle + e^{i\pi/4}|1\rangle)/\sqrt{2}$ using distillation,
(ii)~perform a Bell measurement between the data qubit and the magic state,
(iii)~the measurement outcome, combined with the current Pauli frame, determines whether a corrective S~gate must be applied.
Step~(iii) is a synchronization barrier, where the decoder must provide the relevant frame information before the conditional correction or subsequent adaptive operation can be decided.
Notably, this feedforward requirement is distinct from physically applying every accumulated Pauli correction.
In contrast, most Pauli corrections can still remain in the classical frame.

A key takeaway is that QEC is the substrate through which computation occurs.
Every logical operation translates to a specific number of syndrome rounds, and the decoder must process all of them.
This is why decoder throughput (sustained rounds per second) is as important as single-round latency~\cite{chamberland2026ising}.

\subsection{Where QEC stands today}

The experimental progress of QEC can be organized around several milestones, summarized in Table~\ref{tab:milestones}.

\begin{table}
\centering
\small
\setlength{\tabcolsep}{8pt}
\caption{QEC milestones and industry landscape.}
\label{tab:milestones}
\begin{tabular}{@{}p{2.7cm}p{4.7cm}p{5.2cm}p{2.2cm}@{}}
\toprule
\textbf{Milestone} & \textbf{Meaning} & \textbf{Representative work} & \textbf{Company} \\
\midrule
Below-threshold & Logical error rate decreases exponentially with $d$ & $[[d^2, 1, d]]$ surface code with $d=3/5/7$~\cite{google2025threshold} & Google \\
\addlinespace
Real-time decode integration & Decoder meets latency deadlines in a hardware-integrated loop
 & Union-Find on FPGA, sub-$\mu$s decode~\cite{barber2025realtime} & Riverlane and Rigetti \\
\addlinespace
Logical operations & Logical gates with real-time QEC & $[[7,1,3]]$ color code~\cite{ryan2024high} & Quantinuum \\
\addlinespace
Logical-scale processor & Logical processor with error detection & 48 logical qubits via error-detected $[[8,3,2]]$ code blocks~\cite{bluvstein2024logical} & QuEra \\
\bottomrule
\end{tabular}
\end{table}

The Google result~\cite{google2025threshold} is particularly significant: using a distance-7 surface code on the 105-qubit Willow processor, they demonstrated that logical error rates decrease exponentially with code distance, with an error suppression factor $\Lambda = 2.14 \pm 0.02$.
The logical qubit at $d=7$ surpassed the best physical qubit lifetime on the same device by a factor of $2.4$, crossing the break-even point.
Their primary distance-7 result used offline decoding, without active error correction and logical operations.
Real-time FPGA-integrated decoding was demonstrated by Riverlane and Rigetti~\cite{barber2025realtime} using a Union-Find (UF) decoder with sub-microsecond latency on a Rigetti superconducting processor.
However, their experiment was limited to a memory experiment at small code distances and did not demonstrate logical operations.
Meanwhile, the accuracy of UF is also lower than that of matching-based or neural decoders.

As for logical operations, Quantinuum demonstrated fault-tolerant logical teleportation on the H2 trapped-ion processor~\cite{ryan2024high}, encoding logical qubits in the $[[7,1,3]]$ color code (Steane code) with up to 30 physical qubits.
Crucially, their experiment employed real-time quantum error correction using mid-circuit syndrome measurements and corrections physically applied during the protocol.
Building on this, their most recent work successfully extended real-time QEC to a universal set of fault-tolerant logical operations~\cite{quantinuum2026algorithms}. 
Moving beyond transversal Clifford gates, they implemented highly reliable fault-tolerant non-Clifford operations, realizing logical T-gates with an infidelity as low as $\sim 2.6 \times 10^{-3}$. 
Crucially, this milestone demonstrated that active error correction can practically enhance operational reliability in complex algorithms.
QuEra demonstrated a programmable logical processor based on reconfigurable neutral atom arrays with up to 280 physical qubits~\cite{bluvstein2024logical}.
In particular, a $[[8,3,2]]$ code with $48$ logical qubits is an error-detecting code, where errors are detected and post-selected rather than actively corrected.
Thus, the present experimental landscape contains several distinct milestones: below-threshold memory, hardware-integrated real-time decoding, real-time logical operations on small codes, and error-detected logical processors.
These milestones made important progress towards a complete, scalable FTQC system.

\subsection{Why real-time is now critical}
Three factors make real-time decoding an engineering priority at this stage.

\paragraph{Non-Clifford gates require feedforward.}
Universal quantum computation requires a gate set that includes both Clifford gates and at least one non-Clifford gate, typically the T gate~\cite{nielsen2010quantum}.
As discussed in Section~\ref{subsec:qec_basics}, the Pauli frame must be resolved before each T gate, creating a synchronization barrier.
Between two consecutive T gates, the decoder has a budget of all intervening QEC rounds to drain any accumulated backlog.
If the decoder's sustained throughput cannot keep pace with the syndrome generation rate averaged over this window, backlog accumulates and can force synchronization stalls at feedforward barriers.
In a typical FTQC algorithm, T gates appear frequently:
Gidney and Eker{\aa}~\cite{gidney2021factor} estimate $\sim$$3.7 \times 10^{9}$ Toffoli gates (each requiring several T gates) for RSA-2048 factoring.
The relevant metric is thus whether the decoder can sustain the throughput demanded by the algorithm's T-gate density.

\paragraph{Slow decoding erodes the error threshold at synchronization stalls.}
Decoder lag is not automatically harmful during Clifford-only streaming operation, because corrections can remain in the Pauli frame.
The problem arises when unresolved syndrome data reaches a synchronization barrier and forces the QPU to idle while the decoder catches up.
During such decoder-induced stalls, data qubits continue to accumulate physical errors.
If the resulting effective per-round error rate $p_\text{eff}$ approaches or exceeds the code threshold $p_\text{th}$, error correction becomes counterproductive.
Keeping the end-to-end feedback latency and backlog sufficiently small at synchronization points is therefore a basic requirement for real-time QEC.

\paragraph{Backlog stalls the logical progress.}
Even when the effective error rate remains below threshold, a decoder that cannot keep up on average stalls logical progress.
Because the decoder is allowed to lag during Clifford operations, a temporary backlog is acceptable.
However, if the decoder's sustained throughput $\mu$ is less than the syndrome arrival rate $\lambda$, the backlog grows linearly and eventually causes the logical processor to stall at synchronization barriers associated with T gates or other adaptive operations~\cite{terhal2015online}.
For algorithms dominated by T gates~\cite{gidney2021factor}, each microsecond of decoder-induced stall translates directly to hours of additional runtime.
In the worst case, backlog growth is unbounded, and the logical progress rate degrades to zero.

\subsection{Decoder landscape and open-source ecosystem}

The choice of decoder algorithm is central to real-time QEC system design~\cite{battistel2023realtime}.
Table~\ref{tab:decoders_overview} provides a high-level comparison of the major decoder families; detailed analysis follows in Chapter~\ref{sec:decoders}.

\begin{table}
\centering
\small
\setlength{\tabcolsep}{8pt}
\caption{Overview of major decoder families.}
\label{tab:decoders_overview}
\begin{tabular}{p{3.7cm}p{3.2cm}p{3.2cm}p{3.7cm}}
\toprule
\textbf{Decoder} & \textbf{Complexity} & \textbf{Code support} & \textbf{Hardware} \\
\midrule
Minimum-weight perfect matching (MWPM) & $O(n)$ avg.\ (Sparse Blossom)  & Surface code & CPU~\cite{higgott2025sparse}, FPGA~\cite{microblossom} \\
\addlinespace
Union-Find (UF) & $O(n\,\alpha(n))$ & Surface code & FPGA~\cite{barber2025realtime, das2022afs} \\
\addlinespace
Belief propagation + ordered statistics decoding (BP+OSD) & $O(n^3)$ & CSS code (including qLDPC) & CPU; GPU acceleration \\
\addlinespace
Neural network (NN) & $O(n)$ inference  & Surface / color code & GPU / TPU; hardware-dependent inference~\cite{bausch2024alphaqubit, senior2025alphaqubit2} \\
\bottomrule
\end{tabular}
\end{table}

The open-source ecosystem has matured significantly.
\texttt{Stim}~\cite{gidney2021stim} provides fast stabilizer circuit simulation with circuit-level noise models, generating detection events that can be fed to any decoder.
\texttt{PyMatching}~\cite{higgott2025sparse} implements sparse-blossom MWPM decoding, achieving $\sim$1 million errors per core-second.
Together with libraries such as \texttt{ldpc} for BP and BP+OSD decoding~\cite{roffe2020decoding}, these tools form a practical toolchain for QEC research and benchmarking.
On the proprietary side, NVIDIA has released the CUDA-Q platform for heterogeneous quantum-classical computing that is not fully open-sourced.

Having surveyed the experimental landscape and decoder ecosystem, we now formalize the system model that captures these constraints quantitatively.
\section{System Model for Real-Time QEC}
\label{sec:system_model}

This chapter formalizes the system model for real-time decoding.
Building on Chapter~\ref{sec:background}, we first define the end-to-end data path and the two timing regimes that matter for real-time QEC, then instantiate that model for superconducting, trapped-ion, and neutral-atom platforms, and finally summarize the system-level metrics used throughout the paper.
The total feedback latency is the sum of contributions from stabilizer circuit execution, readout, data transport, decoding, and feedforward control.
An improvement in decoder speed is of limited value if the data link or controller is instead the bottleneck, so the correct perspective is necessarily system-level.

\subsection{End-to-end data path}
\label{sec:data_path}

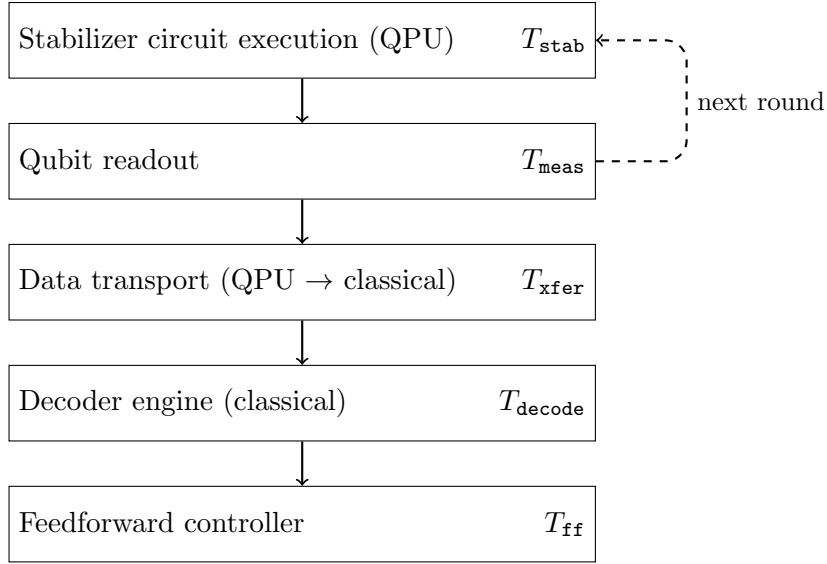
\begin{figure}
\centering
\begin{tikzpicture}[
    node distance=1.6cm,
    block/.style={rectangle, draw, text width=7.5cm, align=center, minimum height=1cm, font=\small},
    arrow/.style={->, thick}
]
\node[block] (stab) {Stabilizer circuit execution (QPU) \hfill {\ttfamily $T_\text{stab}$}};
\node[block, below of=stab] (readout) {Qubit readout \hfill {\ttfamily $T_\text{meas}$}};
\node[block, below of=readout] (transport) {Data transport (QPU $\to$ classical) \hfill {\ttfamily $T_\text{xfer}$}};
\node[block, below of=transport] (decoder) {Decoder engine (classical) \hfill {\ttfamily $T_\text{decode}$}};
\node[block, below of=decoder] (feedback) {Feedforward controller \hfill {\ttfamily $T_\text{ff}$}};

\draw[arrow] (stab) -- (readout);
\draw[arrow] (readout) -- (transport);
\draw[arrow] (transport) -- (decoder);
\draw[arrow] (decoder) -- (feedback);

\draw[arrow, dashed, rounded corners=8pt]
    (readout.east) -- ++(1.2,0) |- node[right, pos=0.25, font=\footnotesize, align=left] {next round} (stab.east);
\end{tikzpicture}
\caption{End-to-end data path from syndrome extraction to classical feedback.
After readout, the QPU can immediately begin the next stabilizer round (dashed arrow) while the classical pipeline processes the syndrome data in parallel.
The timing constraints on the classical stages differ between the two operating regimes described in Section~\ref{sec:two_regimes}.}
\label{fig:critical_path}
\end{figure}

A fault-tolerant quantum computer interleaves quantum and classical processing in a continuous loop.
On the classical side, the syndrome stream traverses several stages, each contributing latency.
We decompose the path into five stages and assign each a latency symbol in Figure~\ref{fig:critical_path}.

Three quantities appear repeatedly in the rest of this chapter.
The first is the QEC round time
\begin{equation}
T_\text{round} \geq T_\text{stab} + T_\text{meas},
\end{equation}
which is lower-bounded by the stabilizer-extraction circuit and readout.
The second is the syndrome arrival rate together with the decoder's sustained throughput
\begin{equation}
\lambda = \frac{1}{T_\text{round}},\qquad\mu \approx \frac{1}{\bar{T}_\text{decode}},
\end{equation}
where $\bar{T}_\text{decode}$ is the mean decode time per round in steady-state streaming operation.
The third is the synchronization latency
\begin{equation}
T_\text{sync}^{(0)} = T_\text{xfer} + T_\text{decode}^{\text{(last)}} + T_\text{ff},
\end{equation}
where $T_\text{xfer}$ is the transport latency, $T_\text{decode}^{\text{(last)}}$ is the decode latency of the final block at a synchronization barrier, and $T_\text{ff}$ is the latency for the frame update and the controller to apply a conditional operation.

The choice of $T_\text{round}$ is constrained by several competing factors.
It must be large enough to accommodate the quantum operations and readout, but small enough that decoherence and idle errors do not dominate.
At the same time, a shorter $T_\text{round}$ increases pressure on the classical stack, whereas a longer $T_\text{round}$ slows the overall algorithm.
Which constraint dominates depends strongly on the hardware platform.

\subsection{Two operating regimes}
\label{sec:two_regimes}

As established in Section~\ref{subsec:qec_basics}, the timing requirement on decoding depends on the type of logical operation being performed~\cite{fowler2012surface, terhal2015online}.
In both regimes, the classical pipeline runs in parallel with the next stabilizer circuit, so the decode stage needs only to keep pace on average and catch up before synchronization points, as we now formalize~\cite{skoric2023parallel}.

\paragraph{Streaming regime (Clifford gates).}
During Clifford-only computation, corrections are tracked in the Pauli frame without immediate physical intervention.
The decoder consumes a continuous stream of syndrome data and produces frame updates, but no logical operation depends on having the corrections applied immediately.
The requirement in this regime is throughput stability: $\mu \geq \lambda$ must hold in steady state.
If $\mu < \lambda$, a backlog $B(t)$ accumulates approximately linearly,
\begin{equation}
    B(t) \approx (\lambda - \mu)\,t.
\end{equation}
Temporary backlog caused by hard-to-decode rounds is acceptable provided that it is drained before the next synchronization barrier.
There is therefore no per-round hard deadline in the streaming regime.

\paragraph{Synchronization regime (non-Clifford gates).}
Before a non-Clifford gate or any other adaptive operation, the Pauli frame must be fully resolved.
At that point the decoder must have processed all relevant syndrome data, including any accumulated backlog.
If it has not, the QPU must idle at the barrier while the decoder catches up, allowing additional physical errors to accumulate.

Let $B_\text{sync}$ denote the number of unprocessed rounds present when a synchronization barrier is reached.
The resulting synchronization latency can be written as
\begin{equation}
    T_\text{sync} = \frac{B_\text{sync}}{\mu} + T_\text{sync}^{(0)},
\end{equation}
where we conservatively assume the decoder drains the backlog at its steady-state throughput $\mu$.
In practice, the drain rate may be higher because no new syndrome data arrives during the catch-up phase.
$T_\text{sync}^{(0)}$ is inherent to every synchronization event, whereas $B_\text{sync}/\mu$ is the avoidable stall created by insufficient throughput or large tail latency.
Between two consecutive synchronization points, the decoder has a budget of all intervening rounds to absorb latency variation.
A real-time stable system must maintain $\mu > \lambda$ with sufficient margin to absorb tail-latency bursts.
These are the components that the system-level metrics in Section~\ref{sec:metrics} aim to control.

\subsection{Platform-dependent instantiations}
Here, we provide platform-specific instantiations of the system model, focusing on the physical origin of $T_\text{round}$ and the dominant real-time constraints.
Throughout the platform-specific discussion below, we use one consistent bookkeeping rule.
$T_\text{stab}$ collects the pre-measurement operations that create the syndrome record: ancilla preparation, routing or allocation into the interaction region, and the entangling stabilizer-extraction gates themselves.
$T_\text{meas}$ collects the measurement chain that turns that quantum state into a usable digital record and prepares the hardware for the next round: detector acquisition, integrated state classification, any protection operation that is part of making the measurement possible, ancilla reset or qubit reuse, loss identification, and any controller overhead that lies on the measurement critical path.

\paragraph{Superconducting qubits.}
For superconducting surface-code experiments, the dominant pressure comes from short coherence times and microsecond QEC cycles.
One can write
\begin{equation}
T_\text{stab}^{\text{sc}} \approx T_\text{prep}^{\rm anc} + T_{\text{gate}},
\qquad T_\text{meas}^{\text{sc}} \approx T_\text{meas}^{\rm anc} + T_\text{reset}^{\rm anc},
\end{equation}
where $T_\text{prep}^{\rm anc}$ is ancilla preparation before syndrome extraction, $T_{\text{gate}}$ is the duration of entangling gates, $T_{\rm meas}^{\rm anc}$ is the ancilla measurement time, and $T_\text{reset}^{\rm anc}$ is active reset after measurement so that the ancilla can be reused in the next round.
Each weight-4 surface-code stabilizer requires four CNOT layers, and at roughly 30--50\,ns per two-qubit gate~\cite{google2025threshold}, this contributes about 120--200\,ns of circuit depth.
Including ancilla preparation, measurement, and active reset yields a full round budget on the order of a microsecond in current superconducting experiments~\cite{google2025threshold,barber2025realtime}.

On superconducting platforms, this quantum time sets a hard lower bound on $T_\text{round}$, while the short coherence time ($T_2 \sim 30$--$100\,\mu$s~\cite{google2025threshold}) sets an upper-level system pressure against making rounds too slow.
The practical operating point is $T_\text{round} \approx 1\,\mu$s, where only $10^1$--$10^2$ rounds fit within a typical coherence window.
This tight budget places the most demanding real-time constraint on the decoder: throughput $\mu \geq 1\,\text{MHz}$ must be sustained with low tail latency.

\paragraph{Trapped ions.}
For trapped-ion systems, the raw decoder-throughput requirement is usually less severe because the physical coherence time is much longer.
The round-time budget can be written schematically as
\begin{equation}
T_{\rm stab}^{\rm ion} \approx T_{\rm prep}^{\rm anc} + T_{\rm gate},
\qquad T_{\rm meas}^{\rm ion} \approx T_{\rm meas}^{\rm anc} + T_{\rm protect} + T_{\rm reset}^{\rm anc} + T_{\rm ctrl}.
\end{equation}
Compared to superconducting platforms, $T_{\rm meas}^{\rm anc}$ includes an integrated state classification process besides the ancilla measurement time, $T_{\rm protect}$ accounts for spectator-qubit protection during mid-circuit measurement, and $T_{\rm ctrl}$ represents classical control and pulse-scheduling overhead embedded in the measurement-and-feedback chain.
In QCCD or modular trapped-ion architectures, shuttling, splitting and merging, and sympathetic recooling contribute additional extraction overhead before the next valid measurement can be taken~\cite{kielpinski2002architecture}.
Therefore, trapped-ion real-time QEC is usually limited by deterministic closed-loop integration.

\paragraph{Neutral atoms.}
For neutral-atom systems, the dominant constraints differ again.
Reconfigurable atom arrays must also route atoms between storage, entangling, and readout zones, and may need to detect atom loss and reload or reuse qubits.
The round-time budget can therefore be written schematically as
\begin{equation}
T_{\rm stab}^{\rm atom} \approx T_{\rm prep}^{\rm anc} + T_{\rm move} + T_{\rm gate},
\qquad T_{\rm meas}^{\rm atom} \approx T_{\rm image} + T_{\rm loss} + T_{\rm reuse} + T_{\rm ctrl},
\end{equation}
where $T_{\rm move}$ is the atom-routing time before syndrome extraction, $T_{\rm image}$ is the state-detection and image-processing time, $T_{\rm loss}$ is the time required to identify atom loss or leakage from the readout record, and $T_{\rm reuse}$ accounts for qubit reuse or reloading after that readout.
The Rydberg-blockade CZ gate takes roughly 0.3--0.5\,$\mu$s~\cite{bluvstein2024logical}, giving approximately 1--2\,$\mu$s of gate depth for a weight-4 stabilizer.
For flexible-layout codes, including possible future qLDPC implementations, the gate depth alone is not the main systems metric.
In this case, atom movement, image processing, loss handling, and reuse policy often dominate the full round budget.
Recent architecture studies summarize atom motion and readout as the dominant bottlenecks, with a combined budget ranging from about $100\,\mu\mathrm{s}$ to several milliseconds depending on whether the implementation is optimized for speed, continuous operation, and integrated reload~\cite{bluvstein2026neutralatomft,cain2026shor}.

On trapped-ion and neutral-atom platforms, coherence times are orders of magnitude longer ($T_2 \sim 1$--$10$\,s~\cite{egan2021fault,bluvstein2024logical}), so decoherence does not limit the number of rounds.
The motivation to minimize $T_\text{round}$ is instead algorithmic.
Slower rounds mean slower logical operations, and algorithm runtime scales linearly with $T_\text{round}$.
For example, Cain et al.~\cite{cain2026shor} estimate that a time-efficient neutral-atom architecture can factor RSA-2048 in about 97 days with roughly $1.0\times10^5$ physical qubits, assuming a 1\,ms round time.
Under this estimation, increasing $T_\text{round}$ by an order of magnitude would push the runtime toward years.
Therefore, the hardware control stack must sustain millisecond-level repeated operation with high duty cycle over very long algorithmic runtimes.
The decoder throughput requirement ($\mu \geq 1\,\text{kHz}$) is accordingly three orders of magnitude more relaxed than for superconducting platforms.
However, these platforms might use qLDPC codes with larger block sizes, and decoders such as BP+OSD exhibit super-linear scaling with block size~\cite{roffe2020decoding}, which partially offsets the slower round rate.
Whether the relaxed throughput budget is sufficient for large-distance qLDPC decoding remains an open question.

Table~\ref{tab:platform_realtime_constraints} summarizes the dominant real-time constraint for each platform.

\begin{table}
\centering
\setlength{\tabcolsep}{8pt}
\small
\begin{tabular}{p{0.25\linewidth}p{0.7\linewidth}}
\hline
\textbf{Platform} & \textbf{Typical dominant real-time constraint} \\
\hline
Superconducting qubits &
Short QEC cycle, short coherence time, stringent tail-latency requirement \\
Trapped ions &
Mid-circuit measurement, ancilla reset, spectator protection, pulse-level feedforward, and possible QCCD shuttling or recooling \\
Neutral atoms &
Atom transport, image-based readout, loss detection, qubit reuse or reloading, and feedforward scheduling \\
\hline
\end{tabular}
\caption[Platform-dependent real-time constraints for QEC]{Platform-dependent real-time constraints for QEC. The same abstract condition $\mu \geq \lambda$ applies to all platforms, but the physical origin of $T_{\rm round}$ and the relevant synchronization bottlenecks differ substantially.}
\label{tab:platform_realtime_constraints}
\end{table}

\subsection{System-level metrics}
\label{sec:metrics}

We define four metrics that collectively characterize real-time readiness.

\begin{enumerate}
    \item \textbf{Sustained throughput} $\mu \geq \lambda$: the classical stack must keep pace with the syndrome stream on average.
    For superconducting platforms ($\lambda \sim 1\,\text{MHz}$), this is a particularly stringent constraint.

    \item \textbf{Tail latency} ($p_{99}$): high-percentile decode latency.
    Tail events cause transient backlog spikes.
    A system where $p_{99} \gg 1/\mu$ may satisfy the throughput requirement on average yet accumulate dangerous backlog before synchronization barriers~\cite{dean2013tail}.

    \item \textbf{Backlog at synchronization} $B_\text{sync}$: the number of unprocessed rounds when a synchronization barrier is reached.
    $B_\text{sync} = 0$ means the avoidable stall vanishes, while $B_\text{sync} > 0$ adds a stall of $B_\text{sync}/\mu$ on top of the fixed pipeline latency.

    \item \textbf{Fixed pipeline overhead} $T_\text{xfer} + T_\text{ff}$: transport and controller response are irreducible overhead at every synchronization point, even when $B_\text{sync} = 0$.
    These terms must be co-optimized with the streaming decoder to minimize $T_\text{sync}$.
\end{enumerate}

The rate at which the system can execute logical operations is ultimately bounded by
\begin{equation}
    f_\text{logical} \leq \frac{1}{R_\text{op} \cdot T_\text{round} + \sum T_\text{sync}},
\end{equation}
where $R_\text{op}$ is the number of QEC rounds consumed by the logical operation, for example $2d$ for lattice surgery.
When the decoder keeps pace ($B_\text{sync} = 0$), $T_\text{sync}$ reduces to $T_\text{sync}^{(0)}$.
If this latency is small relative to $R_\text{op} \cdot T_\text{round}$, the logical rate is limited mainly by the operation's round budget and the hardware round time.
When backlog accumulates, the $B_\text{sync}/\mu$ term in $T_\text{sync}$ dominates, degrading $f_\text{logical}$ and potentially preventing the system from executing non-Clifford gates at the rate required by the algorithm.

\section{Quantum Error-Correcting Codes}
\label{sec:codes}

This chapter surveys the code families relevant to FTQC, with emphasis on properties that affect real-time decoding system design.

\subsection{Surface code}

\paragraph{Stabilizer structure and threshold.}
The rotated surface code~\cite{dennis2002topological, fowler2012surface} is defined on a 2D square lattice with parameters $[[d^2, 1, d]]$ for a $d \times d$ patch.
Each stabilizer involves at most 4 neighboring qubits (weight-4), enabling shallow syndrome extraction circuits compatible with nearest-neighbor hardware connectivity.
The circuit-level depolarizing noise threshold is $\sim$1\%~\cite{fowler2012surface, raussendorf2007fault}, the highest among known topological codes.

\paragraph{Decoder.}
The surface code admits efficient graph-based decoders.
MWPM~\cite{higgott2025sparse} exploits the fact that syndromes can be mapped to a matching problem on a decoding graph.
UF~\cite{delfosse2021almost} achieves near-linear complexity but with lower accuracy.
UF has been implemented on FPGA~\cite{liyanage2023scalable, barber2025realtime, das2022afs}, while MWPM implementations are more commonly reported on CPU~\cite{higgott2025sparse}.

\paragraph{Encoding rate.}
$k/n = 1/d^2 \to 0$: qubit overhead scales quadratically with distance.
A distance-7 code uses 49 physical data qubits for one logical qubit ($\approx$2\% rate).
This poor rate is the primary motivation for exploring qLDPC alternatives.

\paragraph{Logical gates.}
CNOT gate is performed via lattice surgery (merge/split operations)~\cite{horsman2012surface}.
Outcomes from both logical qubits are inputs to the decoder.
T gates are realized through magic state distillation~\cite{bravyi2005universal}, which requires committing the Pauli frame to determine a conditional Clifford correction, as discussed in Chapter~\ref{sec:system_model}.

\subsection{Color code}

\paragraph{Stabilizer structure and threshold.}
Color codes~\cite{bombin2006topological} are defined on trivalent, 3-colorable lattices with parameters $[[d^2, 1, d]]$ (for triangular lattice variants).
Stabilizers are higher-weight (weight-6 on the 6.6.6 lattice, weight-4/8 on the 4.8.8 lattice) compared to the surface code.
The circuit-level noise threshold is about $0.082(3)\%$~\cite{landahl2011fault}, significantly lower than the surface code.

\paragraph{Decoder.}
Decoding is harder than for surface codes: the syndrome structure corresponds to a hypergraph rather than a simple graph, so standard MWPM does not directly apply.
Practical approaches include restriction decoders that project the color code syndrome onto overlapping surface-code-like problems and solve each with MWPM, and Möbius decoders that exploit the code's topological structure.

\paragraph{Encoding rate.}
$k/n = 1/d^2 \to 0$: the same asymptotic scaling as the surface code.
The overhead is comparable in practice.

\paragraph{Logical gates.}
The key advantage of 2D color codes is that the entire Clifford group can be implemented transversally~\cite{bombin2006topological}.
However, the T gate is not transversal in 2D color codes; it requires either magic state distillation (same as surface codes), code switching~\cite{anderson2014fault}, or 3D gauge color codes~\cite{bombin2015gauge} which support a transversal T gate at the cost of significantly more complex stabilizer structure.

\subsection{qLDPC codes}
\label{sec:qldpc}

\paragraph{Stabilizer structure and threshold.}
The bivariate bicycle code~\cite{bravyi2024high} achieves parameters such as $[[144, 12, 12]]$.
Stabilizers are non-local with weight up to 6, requiring long-range connectivity in hardware, making the syndrome extraction circuits deeper than for surface codes.
Bravyi et al.~\cite{bravyi2024high} report a circuit-level noise threshold of $\sim 0.7$--$0.8\%$, comparable to the surface code, though this assumes the degree-6 connectivity is natively available in hardware.

\paragraph{Decoder.}
Standard MWPM does not apply because the Tanner graph is not planar.
A widely used decoder is BP+OSD~\cite{roffe2020decoding}, while BP variants and other specialized qLDPC decoders remain active research directions.
BP+OSD is generally much slower than MWPM or UF for surface-code decoding.

\paragraph{Encoding rate.}
For data qubits, $k/n = \Theta(1)$ is the defining advantage of qLDPC codes.
The $[[144, 12, 12]]$ code achieves an encoding rate of $12/144 \approx 8.3\%$, compared to $1/49 \approx 2\%$ for a $d=7$ surface code.

\paragraph{Logical gates.}
Bravyi et al.~\cite{bravyi2024high} showed that code automorphisms of bivariate bicycle codes enable fold-transversal gates, where physical gates are combined with qubit permutations that preserve the code structure.
These do not generate a universal gate set.
The implementation of lattice-surgery-like protocols for qLDPC codes remains less developed compared to topological codes and constitutes an important open direction.

Table~\ref{tab:code_comparison} summarizes the properties most relevant to system design.

\begin{table}
\small
\setlength{\tabcolsep}{8pt}
\centering
\caption{Code family comparison for real-time QEC system design.}
\label{tab:code_comparison}
\begin{tabular}{@{}p{2.3cm}p{1.4cm}p{2cm}p{2cm}p{5.5cm}@{}}
\toprule
\textbf{Code} & \textbf{Rate} & \textbf{Threshold} & \textbf{Decoder} & \textbf{Logical gates} \\
\midrule
Surface code &
$O(1/d^2)$ &
$\sim$1\% &
MWPM / UF &
Lattice surgery + Magic state distillation \\
Color code &
$O(1/d^2)$ &
$\sim$0.08\% &
Restriction / Möbius &
Transversal Clifford; T via magic-state distillation, code switching, or gauge fixing \\
qLDPC code &
$\Theta(1)$ &
$\sim 0.7$--$0.8\%$ &
BP+OSD / BP variants &
Fold-transversal gates; universal FT gate set remains less mature \\
\bottomrule
\end{tabular}
\end{table}

\subsection{Hardware-code co-design across different platforms}
The choice of code constrains the decoder algorithm, the hardware connectivity, and the system architecture.
Those requirements must then be matched to a concrete hardware stack.

\paragraph{Superconducting qubits.}
Superconducting processors are organized around fixed two-dimensional nearest-neighbor layouts, fast repeated cycles, and highly optimized readout and control pipelines~\cite{google2025threshold,barber2025realtime}.
Therefore, superconducting hardware favors code families whose geometry stays close to the physical wiring graph.
This platform aligns naturally with the surface code, where the stabilizers are local, the extraction circuit is shallow, and the decoder interface is regular enough for streaming implementations on CPU or FPGA backends.
Color codes and qLDPC codes remain relevant as research directions, but on this platform, they immediately translate into extra swap overhead or non-local couplers that must be engineered into the device itself.

\paragraph{Trapped ions.}
In trapped-ion processors, entangling gates are less constrained by a rigid planar graph, so code choices are not tied as tightly to a local two-dimensional layout as in superconducting hardware.
This has made trapped ions a natural testbed for compact logical blocks and protocols that benefit from flexible connectivity.
A ten-qubit QCCD system realized repeated syndrome extraction for the $[[7,1,3]]$ color code with real-time decoding and frame updates~\cite{ryananderson2021realtime}, which was extended to Steane code, Carbon code, and Iceberg code later~\cite{ryan2024high, microsoft2024break, quantinuum2026iceberg}.
These results indicate that trapped ions are especially well matched to small and medium logical blocks and logical primitives that make direct use of flexible connectivity.
That picture changes once the architecture has to scale.
In long chains, motional-mode crowding, calibration drift, and crosstalk become harder to manage.
In QCCD and modular systems, shuttling, splitting and merging, sympathetic recooling, and inter-module communication enter directly into the cost of code execution~\cite{kielpinski2002architecture}.
Therefore, trapped-ion platforms favor codes whose logical benefits are still visible after transport, measurement-zone, and scheduling overheads are included.

\paragraph{Neutral atoms.}
Neutral-atom processors center on reconfigurability.
Optical tweezer arrays can move qubits between storage, interaction, and readout regions, which opens a broad design space for code layouts and logical workflows.
This has already enabled logical-level experiments on hundreds of atoms, including color-code logical qubits, feedforward teleportation, and error-detecting blocks such as $[[8,3,2]]$~\cite{bluvstein2024logical}.
More recent demonstrations combined repeated-QEC ingredients with atom-loss detection and machine-learning decoding on arrays of up to 448 atoms~\cite{bluvstein2026neutralatomft}.
These results motivate interest in higher-rate codes whose non-local stabilizers map more naturally onto reconfigurable arrays than onto fixed nearest-neighbor hardware.
High-rate constructions such as bivariate bicycle codes reduce data-qubit overhead by encoding $k=\Theta(n)$ logical qubits into $n$ data qubits~\cite{bravyi2024high}, becoming strong candidates.
However, atom loss, leakage, or auxiliary-state occupation need to be identified during readout such that the decoder can use location-resolved information instead of treating every fault as an unknown Pauli error~\cite{wu2022erasure}.
For neutral atoms, code design therefore has to be developed together with movement policy, readout layout, loss detection, and the metadata exported to the decoder.

Table~\ref{tab:hardware_code_codesign} summarizes the platform-level code-design implications.
Across all three platforms, the relevant question is which code properties the hardware can support with a manageable extraction circuit, decoder interface, and control schedule.
The practical consequence for a real-time QEC stack is that it must accommodate several code families and preserve enough platform information for the decoder and logical scheduler to act on hardware-specific structure.

\begin{table}
\centering
\small
\caption[Hardware-code co-design considerations]{Hardware-code co-design considerations across major QEC platforms. The table summarizes the code families that are currently most natural on each platform and the system constraints that dominate when scaling them.}
\setlength{\tabcolsep}{8pt}
\label{tab:hardware_code_codesign}
\begin{tabular}{p{0.18\linewidth}p{0.37\linewidth}p{0.37\linewidth}}
\hline
\textbf{Hardware} & \textbf{Code-design opportunity} & \textbf{Main system constraint} \\
\hline
Superconducting qubits &
Surface-code patches with local stabilizers, regular decoder interfaces, and mature lattice-surgery workflows &
Fixed nearest-neighbor geometry, microsecond-cycle timing, and the cost of adding non-local couplers or extra routing depth for heavier-stabilizer codes \\
Trapped ions &
Steane/color-code blocks, compact logical primitives, and protocols that exploit flexible connectivity &
Scaling requires architecture-aware scheduling of motional modes, measurement zones, QCCD shuttling, recooling, and crosstalk management \\
Neutral atoms &
Color-code logical processors, error-detecting blocks, erasure-aware decoding, and possible future high-rate/qLDPC layouts &
Code execution must be co-designed with atom movement, readout zones, loss detection, loss metadata, and real-time scheduling \\
\hline
\end{tabular}
\end{table}

\section{Decoder Algorithm Survey}
\label{sec:decoders}

\subsection{Selection principles}

Decoder selection is not only an algorithm question, but also jointly determined by the code family, the noise model, the deadline constraint, and the target hardware backend (CPU, GPU, or FPGA).
A decoder that is optimal in accuracy may be unusable in a real-time setting if its tail latency is unpredictable.
Moreover, decoder selection is also affected by the available metadata, such as measurement confidence and leakage or erasure flags that are accessible in atomic platforms.

\subsection{Evaluation dimensions}

We evaluate decoders along six axes:
\begin{enumerate}
    \item \textbf{Logical error rate} (accuracy): given physical error rate $p$ and code distance $d$, how low can the logical error rate $p_L$ be driven?
    \item \textbf{Sustained throughput and latency}: decode rounds per second under streaming input as well as tail latency ($p_{99}$).
    \item \textbf{Scalability}: how latency and resource consumption grow with $d$.
    \item \textbf{Hardware mapping efficiency}: suitability for CPU, GPU, or FPGA implementation.
    \item \textbf{Code generality}: surface code only, or applicable to qLDPC and other families.
    \item \textbf{Data adaptability}: ability to use erasure or loss-location information, measurement confidence, and other metadata available in atomic platforms.
\end{enumerate}

\subsection{MWPM}

MWPM~\cite{fowler2015minimum} maps the decoding problem to a matching problem on a graph derived from the syndrome.
It is the gold standard for surface code decoding accuracy.
The classical Blossom~V algorithm has $O(n^3)$ worst-case complexity.
The Sparse Blossom algorithm~\cite{higgott2025sparse}, implemented in \texttt{PyMatching}~v2, exploits syndrome sparsity (most stabilizers report no error) to achieve near-linear average-case performance~\cite{higgott2025sparse}.
At $d \leq 13$, a single CPU core running Sparse Blossom can meet a 1\,$\mu$s deadline for most syndrome configurations at a low physical error rate $p=0.001$ (also see Sec.~\ref{sec:decode_latency}), while rare high-weight error configurations might cause tail latency spikes well beyond the deadline.
At $d \geq 15$, even average decode latency may exceed real-time budgets on CPU implementations.
MWPM is therefore a strong baseline for surface-code accuracy, but its real-time readiness depends on the full latency distribution, the syndrome density, and the hardware path used to deliver detection events to the decoder.

\subsection{UF}

UF decoder~\cite{delfosse2021almost} operates by growing clusters around syndrome defects and merging them using a union-find data structure.
It runs in almost-linear time $O(n \cdot \alpha(n))$, where $\alpha$ is the inverse Ackermann function, achieving a lower and more predictable latency profile than MWPM.
However, its accuracy is below MWPM, with a higher logical error rate near threshold~\cite{delfosse2021almost}.
UF has been implemented on CPUs and FPGAs by multiple groups~\cite{liyanage2023scalable, barber2025realtime}, achieving sub-microsecond decode times for moderate $d$.
The data structure operations, including find with path compression and union by rank, map well to hardware.
Related hardware-efficient surface code decoders in the same spirit include the Collision Clustering decoder~\cite{barber2023collision} and the Local Clustering decoder~\cite{ziad2025lcd} demonstrated on FPGA.

\subsection{BP+OSD}

BP+OSD~\cite{panteleev2021degenerate, roffe2020decoding} is one of the most widely used general-purpose baseline decoders for qLDPC codes, where MWPM and UF are not applicable.
BP iterates message-passing on the Tanner graph of the code, while OSD post-processing corrects residual errors via Gaussian elimination on the most reliable bits.
The message-passing structure is inherently parallel, where each variable node and check node can be updated independently.
This makes BP well-suited to GPU acceleration.
However, convergence on quantum codes is often poor due to short cycles in the Tanner graph, requiring many iterations.
As for the OSD part, it involves Gaussian elimination on an $m \times n$ matrix followed by enumeration of weight-$w$ error patterns.
OSD-0 (Gaussian elimination only, no enumeration) takes $\sim$100\,$\mu$s on a CPU for a $[[144, 12, 12]]$ code~\cite{roffe2020decoding}.
Moreover, higher-order OSD-$w$ adds combinatorial overhead that grows exponentially in $w$.
For superconducting platforms ($\sim$1\,$\mu$s cycle time) this is still prohibitive, but for neutral-atom arrays with $\sim$1\,ms cycle times, BP+OSD may be viable if total latency can be kept within a few hundred microseconds.

\subsection{Neural network decoders}

Neural network decoders~\cite{torlai2017neural, gicev2023scalable} use graph neural networks (GNNs), recurrent transformers, or other architectures trained end-to-end on syndrome--error pairs.
These decoders can learn correlated and device-specific noise structures that graph-based decoders miss, but training cost scales with code size and generalization across noise models is limited.
Beyond real-time decoding, neural decoders trained on hardware data can also serve as offline calibration tools and accuracy benchmarks for classical decoders.
Moreover, neural decoders can incorporate device-specific and analog information, including measurement confidence or other metadata in atomic platforms.

AlphaQubit~\cite{bausch2024alphaqubit} is a transformer-based decoder trained on experimental detection event data, leveraging analog measurement signals to capture noise information.
On Google's distance-5 surface code it achieves $\Lambda = 2.18 \pm 0.09$, exceeding correlated matching ($\Lambda = 2.04 \pm 0.02$)~\cite{bausch2024alphaqubit, google2025threshold}, though its original inference latency ($\sim$24\,$\mu$s/cycle) was far above real-time budgets.
AlphaQubit~2~\cite{senior2025alphaqubit2} achieves per-cycle decode latency below the QEC cycle time (${<}1$\,$\mu$s on TPU/GPU) for the surface code to $d=11$ and the color code to $d=9$.

\subsection{Sliding window decoding}

The decoders above are presented as processing a fixed batch of syndrome rounds.
In practice, syndrome data arrives as a continuous stream, and the decoder must commit corrections incrementally without waiting for the full experiment to finish.
Sliding window decoding~\cite{skoric2023parallel} addresses this by maintaining a window of fixed size along the time axis.
The decoder processes the current window, commits the decoding decision for the oldest rounds, then advances the window forward.
Windowed decoding achieves near-identical logical error rates to batch decoding while bounding the per-step computation to a fixed window size, independent of the total experiment length.
Moreover, it can wrap any inner decoder (MWPM, UF, BP+OSD, or NN) and is essential for scaling real-time decoding to long computations.
\section{Benchmark of Decoders}
\label{sec:benchmark}

\subsection{Benchmark principles}

Our benchmark framework evaluates two groups of metrics for accuracy and efficiency:

\begin{enumerate}
    \item \textbf{Error correction quality}: logical error per shot, threshold.
    \item \textbf{Real-time capability}: streaming throughput, $p_{99}$ latency, scalability with $d$.
\end{enumerate}

\subsection{Methodology and environment}
\label{sec:bench_methodology}

All benchmarks use public softwares on the hardware listed in Table~\ref{tab:hw_env}.
For CPU latency measurements (surface-code MWPM and qLDPC BP+OSD), we use a Linux server with an Intel Xeon Platinum 8358P CPU, and pin the process to a single physical core for stable timing.
GPU measurements with cudaq-qec (Section~\ref{sec:qldpc_bench}) are implemented by an NVIDIA A100 80\,GB GPU, using a batch size of 256 shots per decode call.

\begin{table}
\centering
\small
\setlength{\tabcolsep}{8pt}
\caption{Benchmark hardware and software environment.}
\label{tab:hw_env}
\begin{tabular}{ll}
\toprule
\textbf{Component} & \textbf{Value} \\
\midrule
CPU         & Intel Xeon Platinum 8358P \\
GPU         & NVIDIA A100 80\,GB \\
Stim        & 1.15.0~\cite{gidney2021stim} \\
PyMatching  & 2.3.1~\cite{higgott2025sparse} (Sparse Blossom MWPM) \\
ldpc        & 2.4.1~\cite{roffe_ldpc} (BP+OSD, CPU) \\
cudaq-qec   & 0.6.0~\cite{nvidia_cudaq_qec} (BP+OSD, GPU) \\
\bottomrule
\end{tabular}
\end{table}

We simulate under circuit-level depolarizing noise applied uniformly at rate $p$ across all operations using \texttt{Stim}~\cite{gidney2021stim} (see Table~\ref{tab:noise_models}).
It is the standard for circuit-level threshold analysis~\cite{fowler2012surface, raussendorf2007fault}.
However, this noise model neglects leakage, atom loss, erasure metadata, mid-circuit readout crosstalk, coherent errors, or temporal drift due to unavailable hardware information and limitations of \texttt{Stim}.

\begin{table}
\centering
\small
\setlength{\tabcolsep}{8pt}
\caption{Noise model parameters.}
\label{tab:noise_models}
\begin{tabular}{ll}
\toprule
\textbf{Noise channel} & \textbf{Rate} \\
\midrule
After Clifford depolarization      & $p$ \\
After reset flip probability       & $p$ \\
Before measurement flip probability & $p$ \\
Idle data-qubit depolarization (per round) & $p$ \\
\bottomrule
\end{tabular}
\end{table}

\subsection{Surface code: logical error rate}
\label{sec:surface_ler}

\begin{figure}
    \centering
    \includegraphics[width=0.88\textwidth]{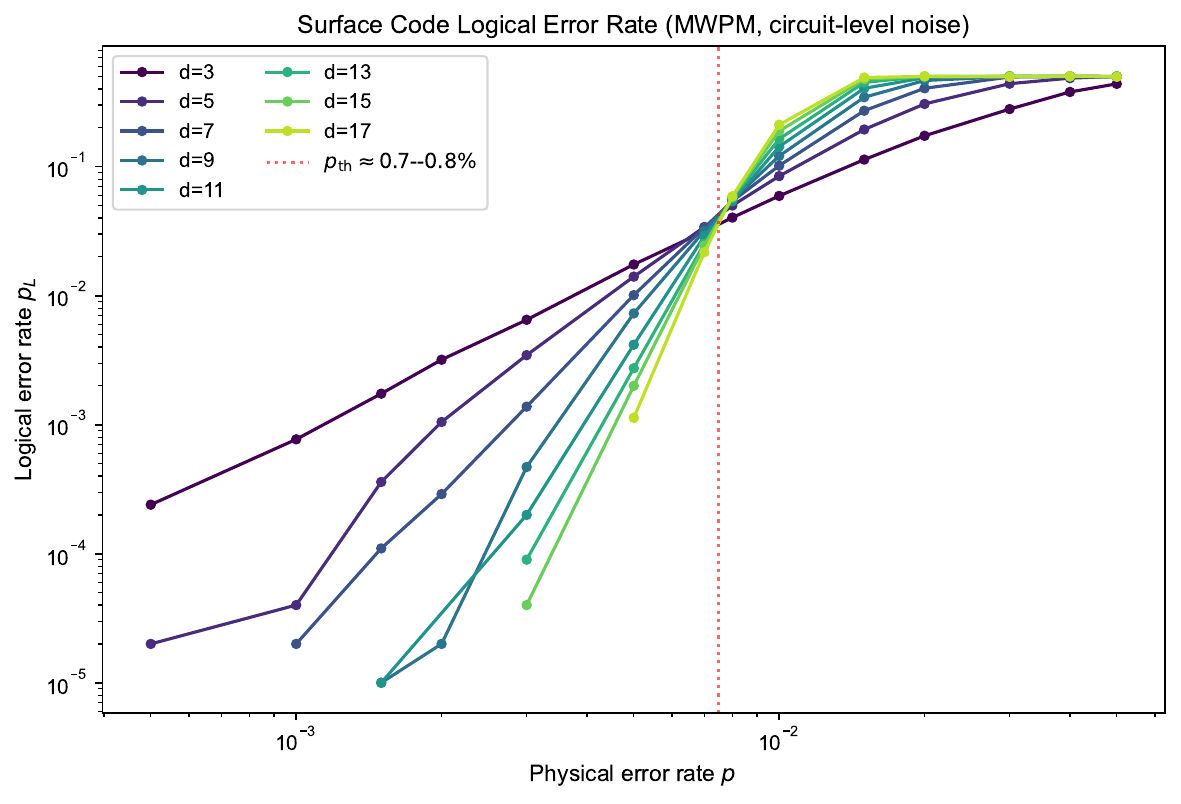}
    \caption{Logical error rate vs.\ physical error rate for rotated surface codes at $d=3$--$17$ using the MWPM decoder.
    Dashed red line: threshold $p_{\rm th} \approx 0.7$--$0.8\%$.}
    \label{fig:logical_error_rate}
\end{figure}

We simulate rotated surface codes at distances $d \in \{3, 5, \cdots, 17\}$ with different physical error rates $p$ from $5 \times 10^{-4}$ to $5 \times 10^{-2}$.
Each configuration uses $d$ rounds of syndrome extraction and $10^5$ shots, sampled via \texttt{Stim}~\cite{gidney2021stim} and decoded with \texttt{PyMatching} v2 (Sparse Blossom MWPM)~\cite{higgott2025sparse}.
Figure~\ref{fig:logical_error_rate} shows the logical error rate as a function of physical error rate, with a vertical dashed line at the observed threshold crossing $p_{\rm th} \approx 0.7$--$0.8\%$.
For $p < 0.007$, increasing $d$ suppresses the logical error rate, while for $p \gtrsim 0.03$, all distances converge toward $p_L \sim 0.5$.
This is consistent with the theoretically predicted circuit-level threshold of $\sim$0.7--1.1\% for the surface code~\cite{fowler2012surface, raussendorf2007fault} under different noise models and decoder variants.
Moreover, $d=17$ achieves a substantially lower $p_L$ than $d=3$ ($\sim$15$\times$ improvement) at $p = 0.005$, demonstrating the practical value of increased code distance even at modest sub-threshold error rates.

\subsection{Batch vs window decoding}
\label{sec:batch_vs_streaming}

In a continuous QEC experiment, the decoder must keep up with the syndrome-extraction rate of one round per $T_{\rm round} \approx 1\,\mu$s in superconducting platforms.
Two decoding strategies partition the syndrome stream differently:

\begin{itemize}
    \item \textbf{Batch}: accumulate $R$ rounds of syndrome data, decode once. 
    The decoder has a budget of $R \times T_{\rm round}$, and dividing by $R$ gives a per-round amortised cost.
    \item \textbf{Window}: decode a $W$-round sliding window and step forward by $C$ rounds for a total of $R$ rounds.
    We choose the $W{=}2d$, $C{=}d$ configuration, which is the standard choice in the sliding window literature~\cite{skoric2023parallel, terhal2015online}.
\end{itemize}

\noindent In both cases, the real-time constraint reduces to the same per-round condition:
$$\frac{\text{Total decoding time}}{R} \leq 1\,\mu\text{s} \quad \text{(per-round amortized cost)}.$$
For slower-cycle platforms, the same metric can be compared against a platform-specific $T_{\rm round}$ defined in Section~\ref{sec:system_model}.
In the following benchmarks, We apply a 50-iteration warmup to eliminate cold-start artifacts, following the methodology of~\cite{chamberland2026ising}.

\subsection{Decode latency: batch and window}
\label{sec:decode_latency}
\begin{figure}
    \centering
    \includegraphics[width=0.88\textwidth]{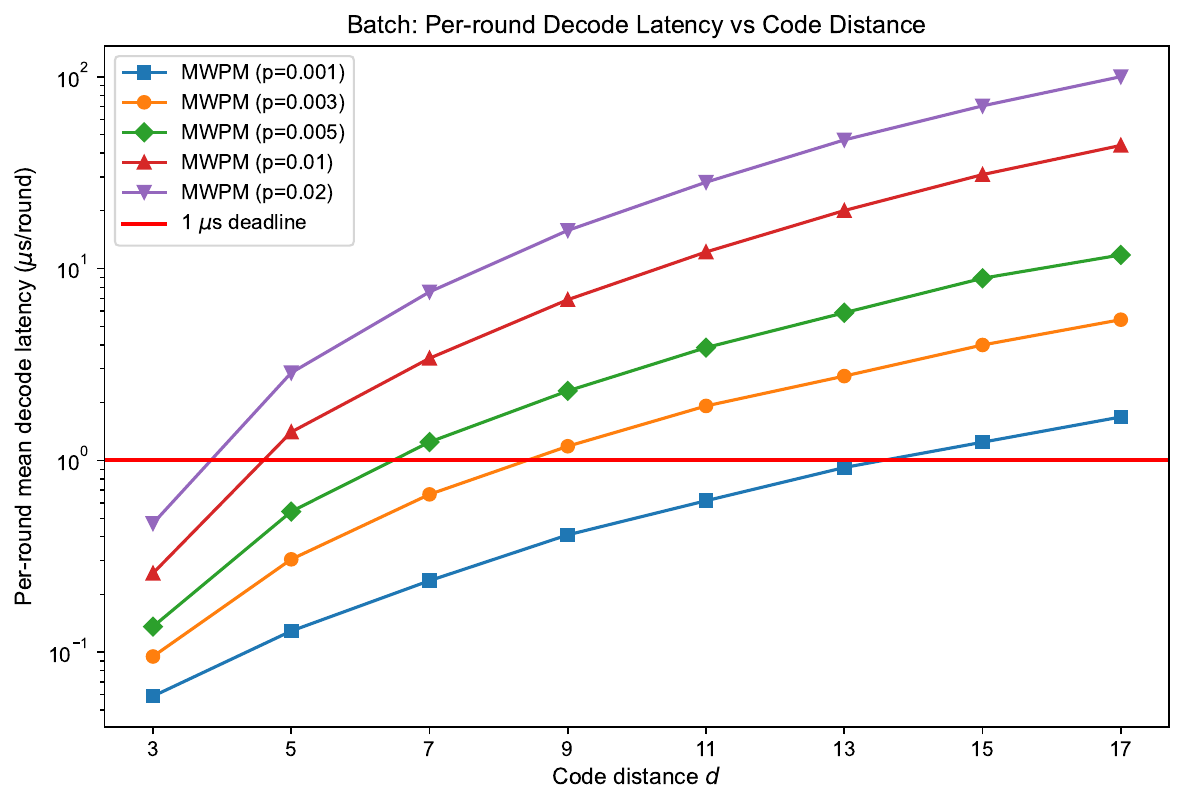}
    \caption{Batch mode: per-round mean decode latency vs.\ code distance at different $p$.
    Each point is the per-round time to decode $R=100$ rounds, averaged over 10\,000 individual experiments on a single CPU core.
    The red line marks the 1\,$\mu$s/round real-time deadline.}
    \label{fig:decode_latency}
\end{figure}

Figure~\ref{fig:decode_latency} shows the mean per-round decode latency in mode at five physical error rates, measured over 10\,000 individual tests using \texttt{PyMatching}.
Each test decodes $R=100$ rounds of syndrome data and the plotted value divides the total decoding time by $R$.
At $p=0.003$, the per-round mean latency ranges from $\sim$0.10\,$\mu$s/round ($d=3$) to $\sim$5.4\,$\mu$s/round ($d=17$), with batch decoding meeting the 1\,$\mu$s deadline at $d \le 7$ (0.10, 0.31, and 0.67\,$\mu$s/round respectively).
Notably, the very recent NVIDIA's Ising-decoder benchmark~\cite{chamberland2026ising} uses $R=8d=104$ syndrome rounds for $d=13$ and $p=0.003$, achieving a per-round cost of $\sim$2.5\,$\mu$s/round on a different CPU, consistent with our measurement of 2.75\,$\mu$s/round under the same noise model and decoder settings.

\begin{figure}
    \centering
    \includegraphics[width=0.88\textwidth]{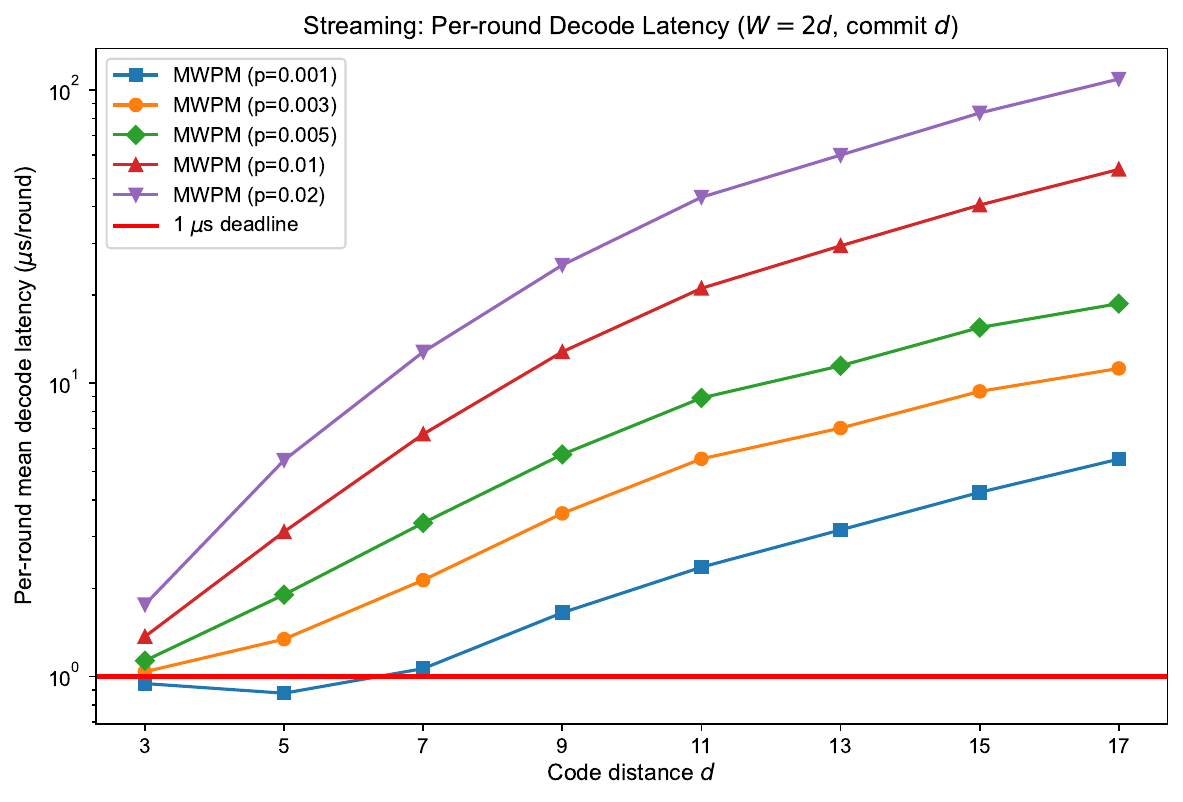}
    \caption{Window mode: per-round mean decode latency vs.\ code distance at different $p$.
    Each point is the total wall-clock time for all sliding-window steps ($W{=}2d$, $C{=}d$, $R{=}100$) divided by $R$, averaged over 10\,000 independent experiments on a single CPU core.
    The red line marks the 1\,$\mu$s/round real-time deadline.}
    \label{fig:streaming_decode_latency}
\end{figure}

\begin{figure}
    \centering
    \includegraphics[width=0.88\textwidth]{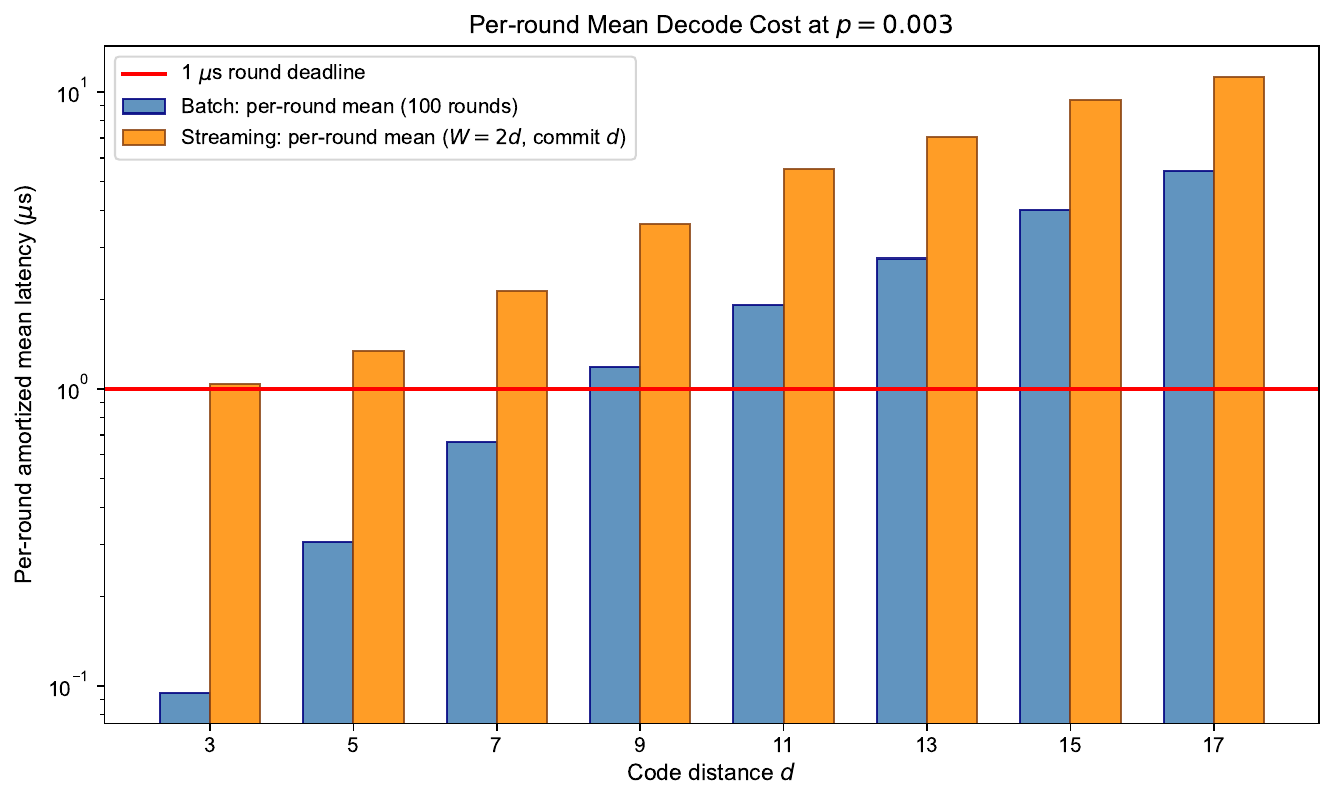}
    \caption{Per-round averaged decode latency at $p=0.003$: batch ($R=100$) vs window ($W{=}2d$, $C{=}d$, $R=100$).
    The red line marks the 1\,$\mu$s deadline.}
    \label{fig:per_round_mean}
\end{figure}

Figure~\ref{fig:streaming_decode_latency} shows the corresponding per-round latency for window mode.
Each point is the per-round cost with $W=2d$, $C=d$, and a total of $R=100$ rounds, averaged over 10\,000 independent experiments.
At $p=0.003$, the window per-round cost ranges from $\sim$1.0\,$\mu$s/round ($d=3$) to $\sim$11.2\,$\mu$s/round ($d=17$), consistently higher than batch because each sliding-window step decodes a $2d$-round window, while committing only $d$ rounds of corrections.
Figure~\ref{fig:per_round_mean} further compares the per-round cost of both modes at $p = 0.003$.
For $d \ge 5$, window decoding is consistently $\sim$2--4$\times$ more expensive than batch.

\subsection{Per-round $p_{99}$ latency: batch vs window}
\label{sec:p99_latency}

Mean latency determines average throughput, but the 99th-percentile tail determines whether the decoder can sustain real-time operation without unbounded backlog accumulation.
A single slow decode event that exceeds the $d$-round budget forces subsequent rounds to queue.
If such events occur frequently enough, the backlog grows monotonically.

\begin{figure}
    \centering
    \includegraphics[width=0.88\textwidth]{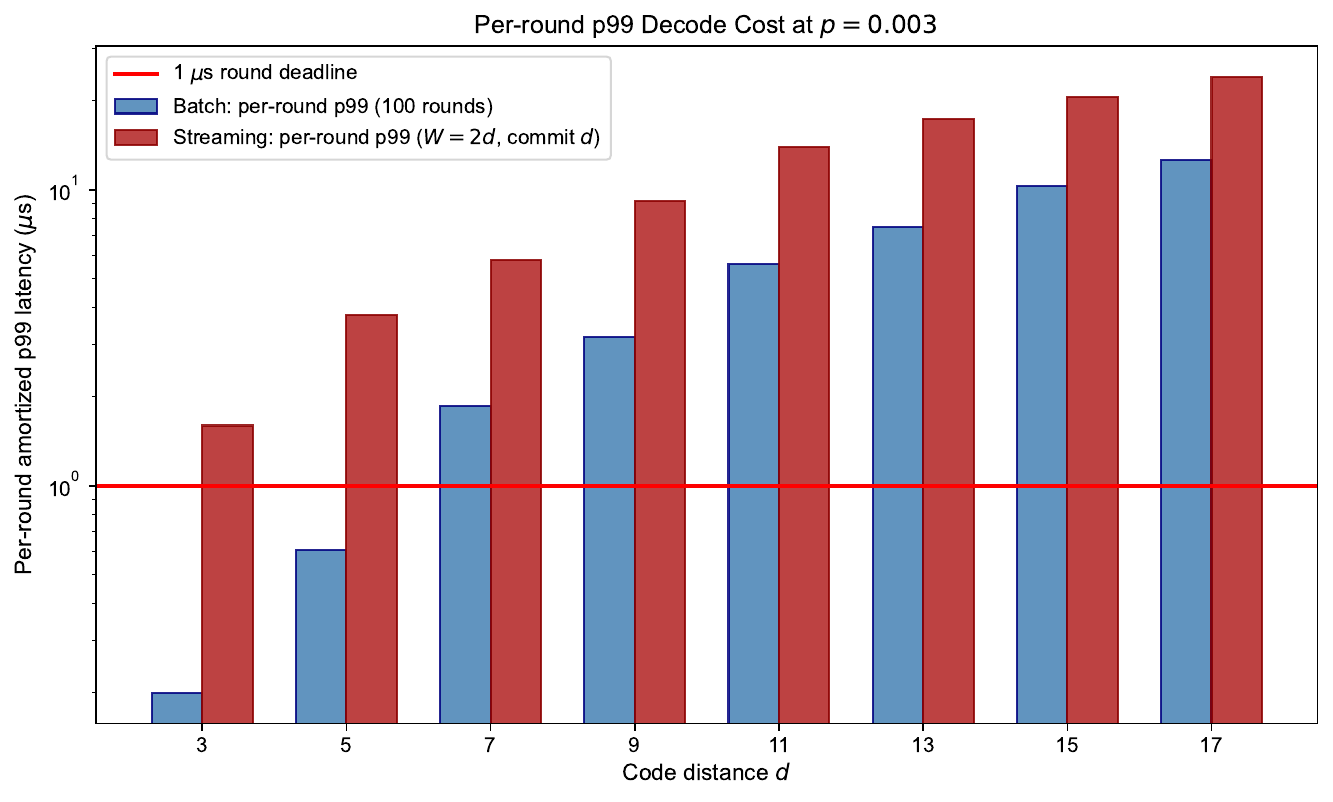}
    \caption{Per-round $p_{99}$ decode latency at $p=0.003$: batch ($R=100$) vs window ($W{=}2d$, $C{=}d$, $R=100$).
    The red line marks the 1\,$\mu$s deadline.}
    \label{fig:per_round_p99}
\end{figure}

Figure~\ref{fig:per_round_p99} compares the per-round $p_{99}$ cost at $p = 0.003$.
Comparing Figures~\ref{fig:per_round_mean} and~\ref{fig:per_round_p99}, the $p_{99}$/mean ratio is $\sim$1.7--3$\times$ for both batch and window (e.g.\ at $d=17$ batch goes from 5.4 to $\sim$12.6\,$\mu$s/round, a $\sim$2.3$\times$ tail), confirming that rare heavy-syndrome configurations produce denser matching graphs and longer decode times.
Concretely, the batch $p_{99}$ per round is 0.20\,$\mu$s at $d{=}3$ and 0.61\,$\mu$s at $d{=}5$, meaning that the $p_{99}$ deadline is met at $d \le 5$.
Meeting $p_{99}$ guarantees at $d \ge 7$ therefore requires either hardware acceleration or algorithmic shortcuts (e.g.\ UF fallback).

The preceding sections establish that a software MWPM decoder running on a single CPU core meets the 1\,$\mu$s per-round deadline only in batch mode at $d \le 7$.
In contrast, at all larger distances and in window mode at every distance the deadline for superconducting platforms is exceeded.
The gap ranges from $1.2\times$ (batch mean at $d = 9$) to $\sim$24$\times$ (window $p_{99}$ at $d = 17$), providing the following implications for system design:
\begin{enumerate}
    \item \textbf{Hardware acceleration is essential.}
    Published FPGA implementations of UF decoder achieve sub-microsecond decode latency for surface codes at moderate distances~\cite{liyanage2023scalable}.
    \item \textbf{Window decoding adds a fundamental overhead.}
    Window decoding is important for long computations, but the larger window ($2d$ rounds) required for correct boundary matching makes window decoding inherently $\sim$2--4$\times$ more expensive than batch.
    \item \textbf{Further hardware and software improvements are essential.}
    Firstly, a longer coherence time may allow a larger deadline (e.g., 10\,$\mu$s) that relaxes the requirements on the decoder.
    Secondly, a smaller error rate (e.g., $p=0.001$) reduces the syndrome density and thus the MWPM computation time, improving batch latency by $\sim$2--4$\times$ compared to $p=0.003$ depending on $d$.
    From the software perspective, algorithmic optimizations can reduce the tail latency.
    For example, a hybrid CNN-MWPM decoder~\cite{chamberland2026ising} can accelerate the decoding for large $d$ and low $p$ by pre-processing the syndrome with a CNN to identify likely error locations, thus reducing the size of the matching graph passed to MWPM.
\end{enumerate}

\subsection{qLDPC benchmark: BP+OSD}
\label{sec:qldpc_bench}

\begin{figure}
    \centering
    \includegraphics[width=0.75\textwidth]{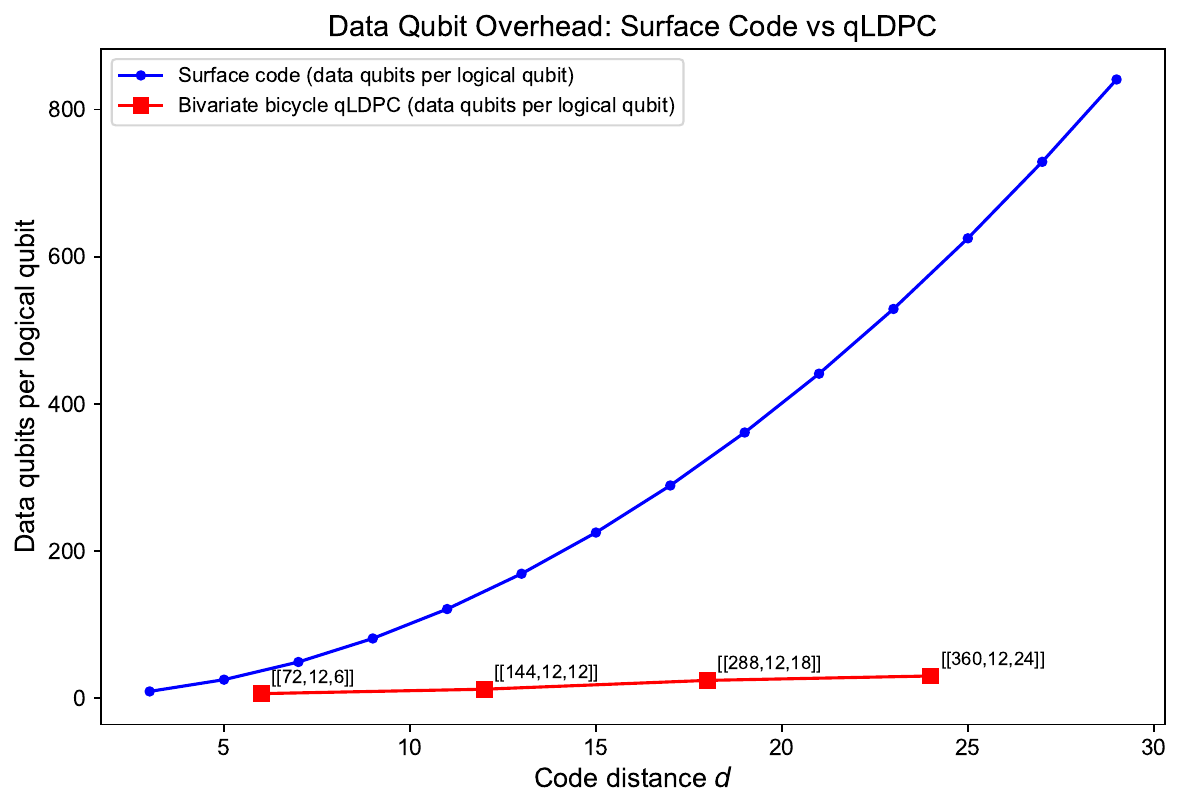}
    \caption{Data qubit overhead per logical qubit: rotated surface code vs.\ bivariate bicycle qLDPC codes~\cite{bravyi2024high} (data qubits only).}
    \label{fig:overhead_comparison}
\end{figure}

Now we switch to consider the qLDPC family of quantum error-correcting codes, which achieve a constant encoding rate $k/n$ and therefore a lower data-qubit overhead per logical qubit than surface-code patches at comparable logical protection.
qLDPC codes are potentially attractive for neutral-atom platforms because reconfigurable arrays may mediate flexible or non-local interaction graphs more naturally than fixed nearest-neighbor architectures.
Figure~\ref{fig:overhead_comparison} compares the data qubit overhead per logical qubit between the surface code and bivariate bicycle qLDPC codes.
This overhead advantage is the primary motivation for qLDPC code research, but comes at the cost of more complex syndrome extraction and decoding.

qLDPC codes are particularly well-suited to neutral-atom platforms (Section~\ref{sec:system_model}), whose syndrome-extraction cycle is currently in the millisecond range.
Therefore, we adopt a \textbf{1\,ms-per-round budget} (corresponding to a $\sim$1\,kHz cycle) as the real-time reference throughout this subsection, replacing the 1\,$\mu$s deadline used for superconducting surface codes above.
For the bivariate bicycle code~\cite{bravyi2024high}, BP+OSD is the standard decoder family in the literature~\cite{panteleev2021degenerate, roffe2020decoding}.
We benchmark two implementations under batch mode with $R=100$ on the same circuit-level depolarizing noise model used above:
\begin{itemize}
    \item \textbf{ldpc}~\cite{roffe_ldpc}: reference C++/Cython implementation, pinned to a single core of an Intel Xeon Platinum 8358P (Linux).
    \item \textbf{cudaq-qec}~\cite{nvidia_cudaq_qec}: batched BP+OSD on a single NVIDIA A100 80\,GB, batch size 256.
\end{itemize}
Both decoders use min-sum BP (max\_iter $=30$, scaling factor $0.75$, parallel schedule) followed by OSD-0.
We sweep three codes ($[[72,12,6]]$, $[[144,12,12]]$, $[[288,12,18]]$) at four physical error rates from $10^{-4}$ to $3\times10^{-3}$.
At $R=100$, $[[288,12,18]]$ is infeasible for \texttt{ldpc} because the OSD reduced-row-echelon step on the space-time decoding matrix exceeds 5\,min per shot, so $[[288]]$ data is reported for \texttt{cudaq-qec} only.
Shot counts are $10\,000$ for $[[72]]$ and $3\,000$ for the two larger codes, where the two decoders agreed on the logical error rate (not shown here) to within statistical noise in every comparable cell.

\begin{figure}
    \centering
    \includegraphics[width=0.88\textwidth]{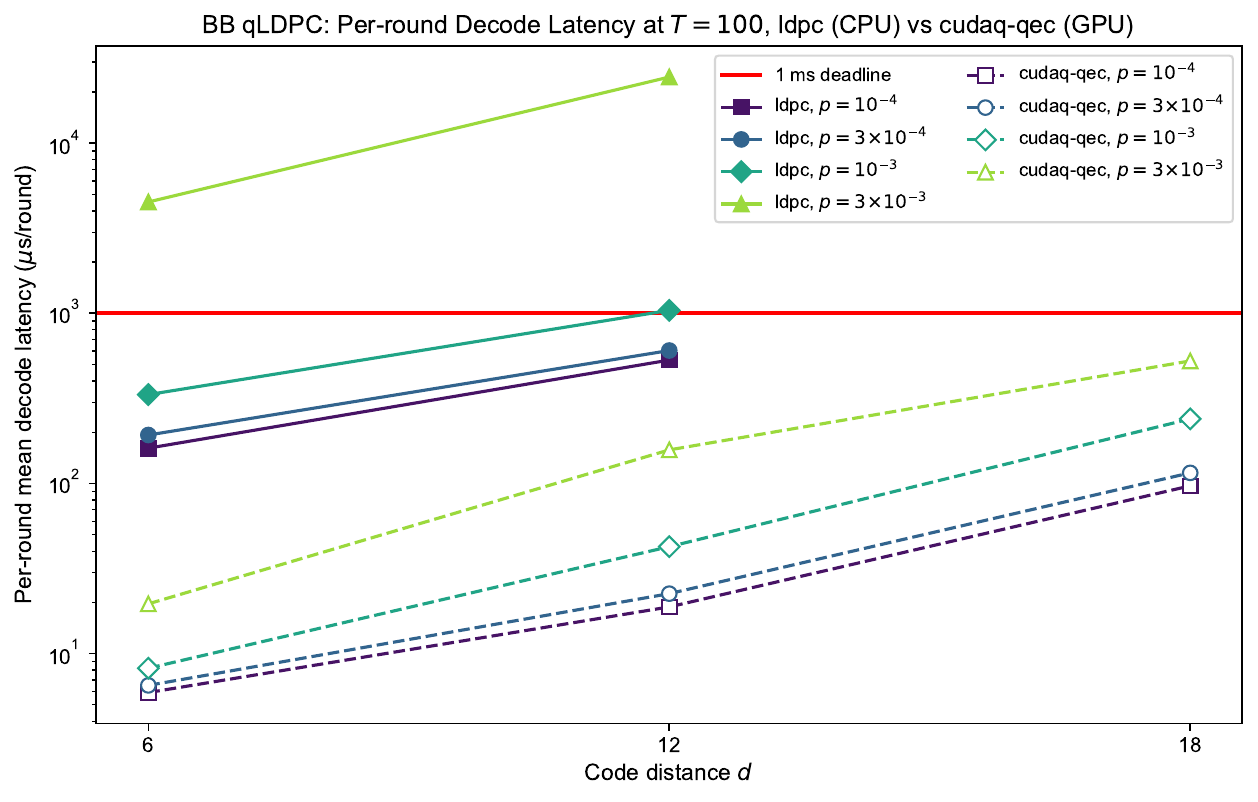}
    \caption{BB qLDPC per-round mean decode latency vs.\ code distance at $R=100$.
    Solid lines: ldpc (single CPU core, BP+OSD-0). Dashed lines: cudaq-qec (A100 GPU, batched BP+OSD-0).
    The red horizontal line marks a 1\,ms reference per round.}
    \label{fig:qldpc_latency}
\end{figure}

Figure~\ref{fig:qldpc_latency} shows the per-round mean decode latency.
At $p=0.001$ the GPU decoder achieves 8.2, 42.5, and 239.5\,$\mu$s/round at $d=6,12,18$ respectively, compared with 332.8\,$\mu$s/round ($d=6$) and 1040.7\,$\mu$s/round ($d=12$) for the CPU decoder.
Three observations follow:
\begin{enumerate}
    \item \textbf{GPU speedup grows with code size and noise.}
    \texttt{cudaq-qec} is $\sim$25--40$\times$ faster than \texttt{ldpc} for $p \leq 0.001$ across both feasible codes, and $\sim$150--230$\times$ faster at $p=0.003$, where OSD on dense matching graphs dominates the CPU runtime.
    \item \textbf{The 1\,ms-per-round target is reachable only on GPU at moderate distances.}
    \texttt{cudaq-qec} meets the 1\,ms-per-round budget for all code distances at all $p$ in the sweep.
    \texttt{ldpc} meets the same budget only for $p \leq 0.001$ at $[[72]]$ (333\,$\mu$s/round) and is already at the edge for $[[144]]$ (1.04\,ms/round).
    At $p=0.003$ it exceeds 1\,ms/round by $4.5\times$ at $[[72]]$ and $24\times$ at $[[144]]$.
    \item \textbf{Scaling differs between CPU and GPU.}
    Moving from $d=6$ to $d=12$ multiplies the per-round latency by $\sim$3.1$\times$ on \texttt{ldpc} but $\sim$5.2$\times$ on \texttt{cudaq-qec} at $p=0.001$, indicating that GPU acceleration is achieved in both stages.
\end{enumerate}

\begin{figure}
    \centering
    \includegraphics[width=0.78\textwidth]{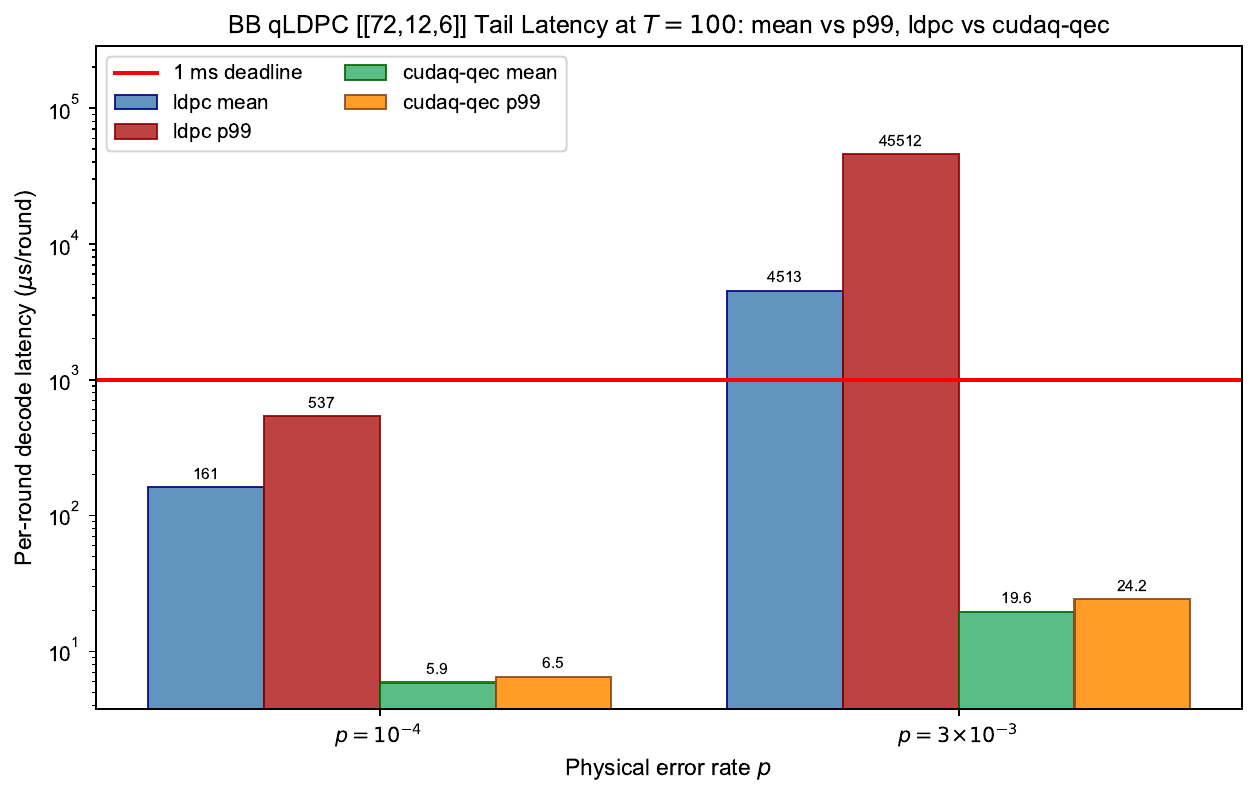}
    \caption{Per-round mean vs.\ $p_{99}$ decode latency for $[[72,12,6]]$ at $R=100$, comparing $\texttt{ldpc}$ (CPU) and $\texttt{cudaq-qec}$ (GPU) at $p=10^{-4}$ and $p=3\times10^{-3}$.
    The red horizontal line marks a 1\,ms reference per round.}
    \label{fig:qldpc_tail_latency}
\end{figure}

Mean latency alone is insufficient for streaming decoding (Section~\ref{sec:p99_latency}).
Figure~\ref{fig:qldpc_tail_latency} compares the per-round mean and $p_{99}$ for $[[72,12,6]]$ at the two error-rate endpoints with full percentile data.
The CPU decoder exhibits an extreme tail: at $p=0.003$ the per-round $p_{99}$ reaches $\sim$45.5\,ms/round, $\sim$10$\times$ the already-high mean of $\sim$4.5\,ms/round, because rare high-weight syndromes trigger long OSD reduced-row-echelon steps.
In contrast, the GPU decoder is essentially tail-free: $p_{99}$/mean $\approx$ 1.1--1.2$\times$ at both error rates, which directly reflects the bounded BP iteration count and the synchronous OSD batch.

The implications for system design parallel those of the surface code:
\begin{enumerate}
    \item BP+OSD on a single CPU core meets a 1\,ms-per-round budget only at low noise and small codes.
    At $p=0.003$ the $p_{99}$ tail exceeds the mean by $\sim$10$\times$ already at $[[72,12,6]]$, making worst-case provisioning impractical.
    \item GPU batching closes most of the gap.
    \texttt{cudaq-qec} sustains $<$1\,ms/round throughput up to $[[288,12,18]]$, and its bounded tail makes it a more credible target for streaming or sliding-window decoding of qLDPC codes.
    \item Approaches that may help include algorithmic alternatives to OSD (e.g.\ closed-branch decoders) and dedicated qLDPC accelerators on FPGA/ASIC, which remains an open direction.
\end{enumerate}

\section{System Architecture and Scaling}
\label{sec:architecture}

The preceding chapters analyzed codes, decoders, and benchmarks for latency and accuracy.
This chapter integrates them into a layered reference architecture for real-time QEC systems and addresses the scaling techniques required to reach the code distances demanded by practical applications.

\subsection{Design principles}

Three principles guide the architecture.
Firstly, each layer owns a single responsibility and exposes a narrow interface, enabling independent evolution.
In particular, a new decoder can be deployed without changing the control electronics, and a new code family can be supported without redesigning the frame manager (concrete data interfaces can be adapted as needed).
Secondly, the dominant operating mode for long-time computing is streaming Pauli frame tracking (Section~\ref{sec:two_regimes}), so the architecture should optimize for sustained throughput in the streaming regime and provide a synchronization mechanism for rare feedforward events.
Thirdly, no single hardware platform is optimal across all decoder--code combinations (Chapter~\ref{sec:decoders}), so the architecture should support CPU, GPU, and FPGA backends behind a common dispatch interface.

\subsection{Six-layer stack}
\label{sec:six_layer_stack}

\begin{figure}
    \centering
    \includegraphics[width=0.9\textwidth]{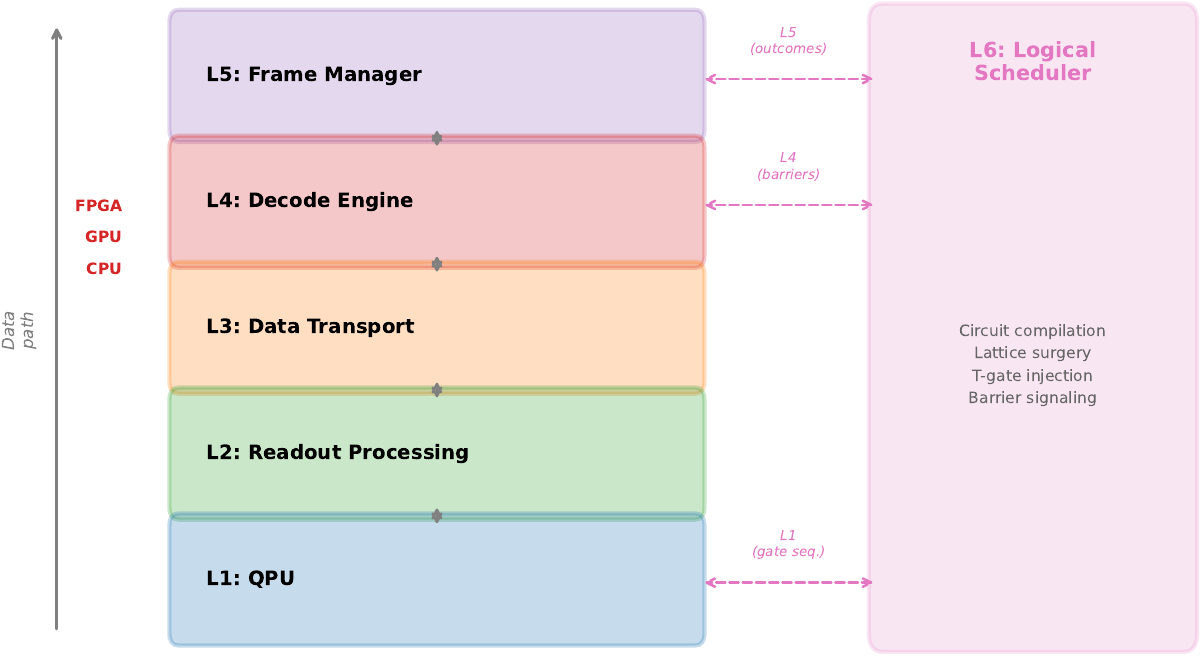}
    \caption{Six-layer reference architecture for real-time QEC.
    Layers~1--5 form the vertical data-path pipeline (solid arrows).
    Layer~6 (Logical Scheduler) is positioned alongside the stack, with lateral interfaces to Layers~1, 4, and~5 (dashed arrows).
    Layer~4 (Decode Engine) is the primary focus of this paper.}
    \label{fig:six_layer_arch}
\end{figure}

We decompose the real-time QEC data path into six layers, mapping directly onto the latency stages defined in Section~\ref{sec:data_path}.
Figure~\ref{fig:six_layer_arch} provides a visual overview of the stack and the data flow between layers.

Layers~1 and~2 (QPU and readout processing) are hardware-specific.
In the simplest surface-code abstraction, they produce a binary syndrome bitstring per QEC round and deliver it to Layer~3, together with some metadata including timestamps, logical-block identifiers, and round indices.
For platforms with image-based readout, leakage detection, or erasure information, the Layer~2 output must be generalized to a structured readout record that includes syndrome bits, confidence information, and erasure/loss flags, as summarized in Table~\ref{tab:stack_structured_readout_record}.

\begin{table}
\centering
\small
\caption{Structured readout record for real-time QEC interfaces.}
\label{tab:stack_structured_readout_record}
\begin{tabular}{p{0.16\linewidth}p{0.30\linewidth}p{0.46\linewidth}}
\hline
\textbf{Field} & \textbf{Meaning} & \textbf{Example use in the system stack} \\
\hline
$s_t$ & Syndrome data & Stabilizer outcomes consumed by the decoder. \\
$c_t$ & Measurement confidence & Photon-count confidence in trapped ions or image-classification confidence in neutral atoms. \\
$e_t$ & Erasure or loss-location flags & Known-location atom loss, leakage, or other detectable erasure events passed to an erasure-aware decoder. \\
$\tau_t$ & Timestamp & Alignment of readout records with QEC rounds, pulse schedules, and decode windows. \\
$b_t$ & Logical-block identifier & Code patch, logical qubit, or logical block label used by the scheduler and frame manager. \\
$r_t$ & QEC round index & Synchronization of streaming decode, backlog accounting, and Pauli-frame updates. \\
\hline
\end{tabular}
\end{table}

Layer~3 (data transport) routes syndrome data from the readout electronics to the decode engine with minimal and deterministic latency.
For cryogenic superconducting systems, this typically involves a high-speed link from the room-temperature control electronics to the decoder host, while a data package with rich information as Table~\ref{tab:stack_structured_readout_record} is required for neutral-atom or trapped-ion systems.
In the Riverlane/Rigetti demonstration~\cite{barber2025realtime}, the decoder is integrated into the quantum processor's gate sequencer FPGA, achieving sub-microsecond end-to-end syndrome transfer by eliminating off-board hops entirely.
When the decoder instead runs on a separate host (e.g., a GPU server), the syndrome packets must traverse a PCIe or network link, adding latency that depends on the interconnect topology and host-side software stack.

Layer~4 (decode engine) receives syndrome packets and other metadata, dispatches them to the appropriate hardware backend, manages the decode queue, and enforces deadlines.
It implements the sliding window protocol to bound per-step computation and enable streaming operation.
When a feedforward barrier is signaled by Layer~5, the engine must drain its backlog and deliver pending corrections within the feedforward deadline.

Layer~5 (frame manager) maintains the Pauli frame, a classical record of accumulated corrections.
In the streaming regime, it appends each decode result to the frame without blocking the pipeline.
At feedforward barriers (e.g., T~gate), it resolves the frame to determine the conditional Clifford correction and signals the QPU control system.
The frame manager also tracks logical measurement outcomes for lattice surgery operations.

Layer~6 (logical scheduler) interfaces with multiple layers simultaneously.
It compiles a logical circuit into a sequence of lattice surgery operations and T~gate injections, programs the QPU control system (Layer~1) with the corresponding physical gate sequences, queries the frame manager (Layer~5) for logical measurement outcomes, and signals the decode engine (Layer~4) with advance notice of upcoming feedforward barriers so that backlog can be drained before hard deadlines arrive.
On trapped-ion and neutral-atom platforms, Layer~6 must also account for platform-specific scheduling constraints, including ion shuttling, recooling, atom movement, readout-zone routing, and loss/erasure handling.

Orthogonal to the layered decomposition, clock synchronization is a cross-cutting concern that spans all layers.
The QPU round counter and decoder timestamps must be aligned to sub-microsecond precision to ensure that syndrome packets, decode results, and control signals refer to the same physical time step.
When the FPGA decoder shares a clock domain with the readout digitizer~\cite{barber2025realtime}, this alignment is inherent.
For systems where the decoder runs on a separate host (e.g., a GPU server), Precision Time Protocol (PTP) or a shared pulse-per-second (PPS) signal can assist in achieving the necessary alignment.

In summary, this six-layer stack characterizes a real-time runtime system for fault-tolerant quantum computation.
In this view, Layers 1--4 form the real-time data plane that continuously converts physical readout into decoded correction information, and Layers 5--6 form the control plane that manages Pauli-frame resolution, logical scheduling, and feedforward decisions.
This runtime-OS perspective makes explicit that scalable QEC requires deterministic coordination among measurement, data transport, decoding, frame management, and logical scheduling.

\subsection{Platform-specific stack implementation}

The six-layer stack in Section~\ref{sec:six_layer_stack} is hardware-agnostic, where the same logical interfaces can describe superconducting, trapped-ion, and neutral-atom systems.
However, the physical implementation of each layer differs substantially across platforms.
These differences should be exposed to the system software to allow the decoder and logical scheduler to optimize for platform-specific constraints.
Since the reference stack introduced above is already closest to a superconducting surface-code implementation, we will focus on the trapped-ion and neutral-atom cases here.

\subsubsection{Trapped ions}

In trapped-ion QEC, the Layer~1 QPU is built from laser- or microwave-driven gates, state-dependent fluorescence readout, and shared motional modes.
Small and intermediate trapped-ion systems benefit from flexible connectivity, which has enabled real-time QEC with the $[[7,1,3]]$ color code and fault-tolerant logical primitives with Bacon--Shor and Steane/color-code blocks~\cite{ryananderson2021realtime,egan2021fault,ryan2024high}.
At larger scale, however, the QPU layer cannot be represented as a uniform all-to-all device.
The logical scheduler must account for mode spectrum crowding, laser-beam addressing constraints, spectator-qubit errors, and the availability of measurement and cooling zones.

Layer~2 is especially important for trapped ions because mid-circuit measurement is not a passive readout operation.
Ancilla fluorescence must be collected while protecting data ions from resonant light scattering, off-resonant excitation, and detection-induced crosstalk.
The readout-processing layer should therefore output not only a binary measurement result but also measurement confidence, leakage flags when available, and a timestamp that can be correlated with the pulse schedule.

Layer~3 routes measurement results from the detection hardware to the decoder and frame manager.
In small systems this path can be implemented through FPGA or real-time CPU control electronics.
In larger QCCD or modular systems, the transport layer must also preserve the association between the readout result and the physical zone, ion identity, logical block, and QEC round.
This is necessary because shuttling and reordering operations can change the mapping between physical ions and logical qubit labels.

Layer~4, the decode engine, usually faces a less stringent raw throughput requirement than in superconducting surface-code systems, because trapped-ion gate and measurement cycles are slower and coherence times are longer.
Nevertheless, deterministic latency remains essential.~\cite{ryananderson2021realtime}.

Layer~5 maintains the Pauli frame and logical measurement outcomes.
For trapped ions, the frame manager must interface with pulse-level control.
Some corrections can remain as software frame updates, while others may require physically applied gates or changes to subsequent pulse phases, which should be represented explicitly in the frame manager.

Layer~6, the logical scheduler, is one of the most platform-specific layers for large trapped-ion systems.
It must compile logical operations into not only gate sequences, but also ion transport, ancilla allocation, measurement-zone usage, recooling, and crosstalk-avoidance constraints.

\subsubsection{Neutral atoms}

Neutral-atom QEC has a different system profile.
The Layer~1 QPU is typically an optical tweezer array with Rydberg-mediated entangling gates and dynamic atom rearrangement.
Atoms can be routed between storage, entangling, readout, and reservoir zones.
A stack in which array reconfiguration, readout metadata, and decoder feedback are tightly integrated is therefore essential to exploit the platform's unique capabilities.

Layer~2 for neutral atoms is often image-based.
The raw data are camera images or site-resolved fluorescence records.
Readout processing must therefore perform state classification, atom-presence detection, and confidence estimation.
Atom loss becomes erasure information that can be used by the decoder.
This is particularly important for alkaline-earth Rydberg atom proposals, where a large fraction of dominant physical errors can in principle be converted into known-location erasures~\cite{wu2022erasure}.
The Layer~2 output should therefore be a structured record containing syndrome bits, erasure flags, confidence scores, and site identifiers.

Layer~3 transports this structured readout record to the decoding and scheduling layers.
For neutral atoms, the data volume and latency profile can differ substantially from a simple bitstream, requiring a more complex transport interface.

Layer~4 should support erasure-aware and geometry-aware decoding.
A neutral-atom decoder may need to combine Pauli-syndrome information with atom-loss locations and measurement confidence.
In addition, neutral atoms are a potential platform for flexible-layout or high-rate codes, which require decoders that understand non-local stabilizers, dynamic connectivity, and possibly time-dependent code layouts.

Layer~5 must maintain not only the Pauli frame but also an erasure and occupancy record.
For neutral atoms, the frame manager should track which physical sites remain occupied, which losses have been declared, and which atoms have been reused.
These records are needed to determine whether a logical operation can continue or whether the logical scheduler should request atom replacement from a reservoir region.

Layer~6 is responsible for array-level scheduling.
It must compile logical operations into Rydberg-gate layers, atom movements, readout-zone routing, reservoir usage, and feedforward actions.
Unlike a fixed-grid architecture, a neutral-atom processor can change the interaction graph during computation.
This flexibility is powerful, but it also means that QEC scheduling is inseparable from physical layout planning.

\subsection{Decode engine internals}

The decode engine (Layer~4) is the most complex layer and merits further decomposition into three subsystems: dispatch, backlog management, and deadline enforcement.

The dispatch policy selects a backend for each syndrome packet based on the code family, current backlog depth, and proximity of the next feedforward barrier.
For surface codes at small $d$ or low $p$, a CPU running MWPM provides sufficient accuracy within the time budget.
At larger $d$ or higher $p$, the dispatcher switches to an FPGA running UF to meet tighter deadlines and demonstrate the QEC loop (though it might not support FTQC due to larger logical error rates).
Runtime conditions can further refine this policy: monitoring the syndrome density (fraction of triggered stabilizers) per round allows adaptive switching between fast and accurate paths, with the threshold calibrated offline from the noise model.

Backlog management monitors the decode queue depth $B(t)$ continuously.
When $B(t)$ exceeds a configurable threshold, the engine escalates by batching multiple rounds into a single decode call to amortize overhead, switching to a faster decoder, or alerting the logical scheduler to pause logical operations until the backlog drains.
This directly addresses the compounding consequences of falling behind identified in Section~\ref{subsec:thesis} (Claim~3): unchecked backlog growth can force synchronization stalls, allowing additional idle errors to accumulate and delaying logical operations at feedforward barriers.

In addition to static decoder selection, the decode engine should also maintain an online policy layer that selects among accuracy-optimized, latency-optimized, and fallback decoders according to real-time system state.
When the backlog is small and the next synchronization point is far away, the system may prefer a high-accuracy decoder such as MWPM or a neural decoder.
When backlog approaches a hard deadline (synchronization regime as defined in Section~\ref{sec:two_regimes}), the engine falls back to a fast approximate decoder (e.g., UF instead of MWPM) and flags the reduced confidence to the frame manager.
Soft deadline misses are logged for monitoring, but do not stall the pipeline.

We additionally propose speculative decoding to run a fast decoder (e.g., FPGA-UF) and an accurate decoder (e.g., CPU-MWPM) in parallel on the same syndrome data.
The fast decoder's result is accepted immediately to meet the streaming deadline, so no stall occurs.
Meanwhile, the accurate decoder runs concurrently.
If it completes before the next feedforward barrier and disagrees with the fast result, the frame manager retroactively applies the more accurate correction.
At low physical error rates ($p \ll p_\text{th}$), the two decoders agree on most of rounds, making retroactive corrections rare.
This strategy bounds worst-case latency at the fast decoder's speed while preserving the accurate decoder's error-correction quality in most cases.

\subsection{Scaling to large code distances}
\label{sec:scaling}

The benchmarks of Chapter~\ref{sec:benchmark} show that even optimized software decoders fail to meet real-time deadlines on a single CPU core beyond moderate code distances.
Scaling to $d \geq 17$ as required for practical algorithms~\cite{gidney2021factor} demands a combination of hardware acceleration and algorithmic decomposition.

\paragraph{FPGA acceleration for surface codes.}
The regular structure of the surface code matching graph maps efficiently to FPGA fabric.
Published implementations achieve sub-microsecond decode latency for UF at $d \leq 11$~\cite{liyanage2023scalable, barber2025realtime}.
Scaling to $d = 17$+ requires either larger FPGAs with sufficient on-chip memory for the matching graph ($O(d^3)$ nodes in the space-time graph) or multi-FPGA partitioning strategies that split the decoding graph along spatial boundaries and reconcile the results.

\paragraph{GPU acceleration for BP+OSD.}
The message-passing phase of BP is embarrassingly parallel: each variable and check node update is independent, making it well-suited to GPU execution.
GPU-parallelized BP has recently reduced decoding latency to below 50\,$\mu$s for $[[784, 24, 24]]$ codes on commodity hardware~\cite{ferraz2025gpubp}.
In practice, BP and OSD can also be pipelined, with the OSD stage on one batch overlapping the BP stage on the next.
Alternatively, Relay-BP serves as an iterative decoding strategy that enhances BP and often reduces or eliminates the need for expensive OSD post-processing~\cite{muller2025improvedbeliefpropagationsufficient}.
FPGA-accelerated Relay-BP has also been demonstrated for qLDPC codes: Maurer et al.~\cite{maurer2025grosscode} implement the Relay-BP algorithm on FPGA for the $[[144,12,12]]$ bivariate bicycle code, achieving an average per-cycle decoding time below 1\,$\mu$s at $p \leq 0.003$.

\paragraph{Parallel window decoding.}
Sliding window decoding is inherently sequential, where each window of $C$ committed rounds must complete within the corresponding data-acquisition interval or syndrome data accumulates.
Skoric et al.~\cite{skoric2023parallel} address this bottleneck with a two-layer parallel schedule: layer~A decodes non-overlapping commit regions in parallel (with buffered windows, typically $W=3C$), and layer~B resolves the gaps using the boundary information output by layer~A.
With this construction, logical fidelity remains comparable to non-windowed decoding in their numerics, while throughput scales near-linearly with process count in the low-overhead regime.
In particular, they report more than one order-of-magnitude speedup at $N_\mathrm{par}=16$~\cite{skoric2023parallel}.

\paragraph{Distributed decoder architecture.}
At large logical-qubit counts, real-time decoding should be treated as a distributed-systems problem~\cite{microblossom}.
A monolithic decoder does not scale indefinitely with the number of logical patches, lattice-surgery boundaries, and concurrent syndrome streams.
The architecture should therefore support decoder sharding, where each patch (or small patch group) is handled by a local decoding worker, while boundary events are coordinated through a dedicated synchronization layer.
In the long term, the dispatcher must concurrently serve both code families on a heterogeneous backend.
\section{From Decoding to Applications}
\label{sec:applications}

The architecture and scaling techniques of the preceding chapter address how to build a real-time decoder system.
This chapter addresses why it matters: how decoder performance propagates through the logical computing stack to determine algorithm viability, and where the critical gaps in the current ecosystem lie.

\subsection{The T-gate bottleneck}

As established in Section~\ref{sec:two_regimes}, Clifford gates operate in the streaming regime and place no feedforward demands on the decoder.
Non-Clifford gates, especially T~gates, are the binding constraint.
Each T gate requires the Pauli frame to be fully resolved before a conditional S correction can be applied, triggering a hard-deadline feedforward event.
Notably, most Pauli corrections can still remain in the classical frame and do not need to be physically applied.

T~gates cannot be implemented transversally on the surface code and instead require magic state distillation~\cite{bravyi2005universal}, where many noisy $|T\rangle$ states are prepared and distilled into a high-fidelity one through a QEC-protected protocol.
A standard 15-to-1 distillation protocol at distance $d$ occupies $\sim$$15 d^2$ physical qubits and runs for $\sim$$5d$ QEC rounds~\cite{litinski2019magic}.
Each distillation cycle consists of preparing 15 noisy $|T\rangle$ ancillas, performing a sequence of CNOT gates and syndrome measurements, and finally measuring the output state to accept or reject it.
The T~factory's output rate is therefore coupled to the decoder's ability to keep up with the syndrome stream and to resolve the required frame information at synchronization points.

The impact on algorithm runtime follows a simple model.
In some resource estimates, the Toffoli gate count serves as an algorithm-level non-Clifford resource metric.
Let $N_\text {Toff}$ be the total Toffoli-gate count and $T_\text{cycle}$ the time per Toffoli gate, limited by the distillation cycle (which is in turn limited by decode + feedforward latency).
The Toffoli-gate-limited runtime is $T_\text{algo} \geq N_\text{Toff} \cdot T_\text{cycle}$.
We illustrate with RSA-2048 factoring, which requires $\sim$$3.7 \times 10^{9}$ Toffoli gates~\cite{gidney2021factor}.
Each Toffoli is implemented by consuming one Toffoli magic state from a dedicated distillation factory~\cite{litinski2019magic}, which is more demanding than T~gate implementation.
As a baseline, we assume each Toffoli factory run also takes $\sim$$5d = 85$ QEC rounds at $d = 17$, giving $T_\text{cycle} \approx 85\,\mu$s per Toffoli at $T_\text{round} = 1\,\mu$s.
With a single Toffoli factory, the serial runtime is $T_\text{algo} \geq 3.7 \times 10^{9} \times 85\,\mu\text{s} \approx 3.6$~days.
Critically, the decoder must sustain this throughput without accumulating backlog over the entire multi-day runtime.
Notably, this is an oversimplified estimate that assumes perfect parallelization of the Clifford gates and no additional overhead from logical qubit routing, but it serves to illustrate the scale of the decoder performance required for a real-world application.

This example confirms that decoder latency is a first-order determinant of computational capability (Claim~3).
The impact is twofold: decoder-induced synchronization stalls allow additional physical errors to accumulate during QPU idle time, and they inflate wall-clock runtime by stalling the logical operations at feedforward barriers.
For near-term demonstrations at small code distances, the primary concern is meeting the per-round deadline set by physical qubit coherence times, where a single missed deadline can invalidate the entire QEC cycle.
For long-running algorithms at higher code distances, the compounding effect dominates.
Even a small sustained throughput deficit accumulates into hours of additional runtime.

\subsection{The middleware gap}

Despite rapid hardware progress, the QEC middleware layer, including decoders, frame managers, control system integration, and benchmarking infrastructure, remains underdeveloped relative to both the hardware below and the algorithms above.
This gap is the central obstacle to scaling from laboratory demonstrations to production FTQC (Claim~1).
We identify three specific shortcomings.

\paragraph{Benchmarking.}
There is no community-accepted benchmark suite for real-time decoder evaluation.
Most papers report only average decode time on a single code distance, omitting the tail latency, sustained throughput, and backlog behavior metrics that matter for real-time operation (Claim~4).
The benchmark framework presented in Chapter~\ref{sec:benchmark} is a step toward filling this gap.
However, due to the limitation of hardware access, it currently relies on software simulations of surface codes and qLDPC codes under a circuit-level noise model.
A more comprehensive noise model is needed to capture the full range of hardware-specific effects, as summarized in Table~\ref{tab:benchmark_scope}.

\begin{table}
\centering
\small
\caption[Scope of the present benchmark noise model]{Scope of the present benchmark noise model. The current simulations are useful for decoder-centric comparison, but they do not include several platform-specific effects that enter a full real-time QEC stack.}
\label{tab:benchmark_scope}
\begin{tabular}{p{0.42\linewidth}p{0.16\linewidth}p{0.34\linewidth}}
\toprule
\textbf{Noise or hardware effect} & \textbf{Included?} & \textbf{Comment} \\
\midrule
Circuit-level depolarizing noise & Yes &
Used as the main benchmark noise model. \\
Reset and measurement bit-flip errors & Yes &
Included through the circuit-level noise model. \\
Idle data-qubit depolarization & Yes &
Included as a per-round idle error channel. \\
Leakage & No &
Not modeled in the current benchmark. \\
Atom loss / erasure & No &
Not modeled; relevant for neutral-atom and erasure-aware decoding studies. \\
Mid-circuit readout crosstalk & No &
Not modeled; relevant for trapped-ion and other mid-circuit measurement platforms. \\
Coherent errors and temporal drift & No &
Not modeled; important for real-device closed-loop validation. \\
\bottomrule
\end{tabular}
\end{table}

\paragraph{Interface standardization.}
The interface between QPU control systems and decoder engines remains ad hoc and vendor-specific.
Recent industrial efforts in this direction, such as NVIDIA's NVQLink, provide a useful case study of interface standardization for real-time QEC~\cite{nvidia_nvqlink}.
In this architecture, syndrome data are streamed from a quantum controller to a GPU-visible buffer, dispatched to a registered CUDA-Q QEC decoder through a low-latency runtime path, and corrections are returned to the controller through the same real-time data plane.
The controller, transport layer, decoder plugin, and feedback path communicate through a common runtime interface.
Nevertheless, it is not a community-governed QEC standard, and till now no open standard exists for syndrome packet formats, correction protocols, or decoder APIs, forcing each hardware--decoder integration to be built from scratch.
Moreover, for atomic platforms, this interface should go beyond binary syndrome packets and include measurement confidence and erasure or loss-location flags.

\paragraph{qLDPC decoder maturity.}
While surface code decoders have matured to meet real-time deadlines at small code distances, qLDPC decoders remain less mature from a real-time systems perspective.
BP+OSD is a widely used decoder for qLDPC codes, but faster specialized decoders, BP variants, and hardware-accelerated implementations are active research directions.

These gaps represent the highest-leverage opportunities for new entrants.
They are critical for scaling QEC but require systems engineering expertise instead of quantum physics breakthroughs.
Real-time QEC naturally intersects with deterministic networking, precision timing, heterogeneous computing, and large-scale telemetry.
These are areas where telecom and cloud-infrastructure expertise can contribute directly to FTQC system engineering.
In particular, low-latency data transport, PTP/PPS-based clock synchronization, FPGA/GPU resource orchestration, distributed control-plane reliability, and continuous runtime telemetry require the ability to engineer a reliable, low-latency, distributed control infrastructure around the quantum processor.
\section{Conclusion}
\label{sec:conclusion}

This paper has examined real-time quantum error correction from a systematic engineering perspective, moving beyond algorithmic benchmarks to address the full data path from syndrome extraction to logical operations.
We conclude by revisiting the six claims of our thesis and summarizing the evidence presented for each.

\paragraph*{Claim~1: The bottleneck has shifted from algorithms to systems.}
Our benchmarks confirm that decoder algorithms such as MWPM can meet real-time deadlines at small code distances on a single CPU core.
The remaining challenges include sustained throughput, bounded tail latency, backlog management, and end-to-end integration, all of which are systems engineering problems (Chapter~\ref{sec:system_model}).

\paragraph*{Claim~2: Most closed loops do not involve active physical correction.}
We formalized the distinction between streaming and synchronization regimes (Section~\ref{sec:two_regimes}) and showed that hard deadlines arise at synchronization barriers, where the relevant Pauli-frame information must be available for adaptive measurements, conditional Clifford corrections, magic-state consumption, or other feedforward decisions.
This distinction is essential for correctly specifying decoder requirements.

\paragraph*{Claim~3: A decoder that falls behind has compounding consequences.}
The synchronization latency formula $T_\text{sync} = B_\text{sync}/\mu + T_\text{sync}^{(0)}$ (Section~\ref{sec:two_regimes}) shows that accumulated backlog accumulates physical errors and inflates logical operation latency.
The RSA-2048 example (Section~\ref{sec:applications}) demonstrates that even small throughput deficits translate to days of additional runtime.

\paragraph*{Claim~4: Tail latency is the operationally relevant metric.}
Our $p_{99}$ latency measurements (Section~\ref{sec:p99_latency}) show a $\sim$1.7--3$\times$ ratio over mean latency for MWPM, confirming that rare heavy-syndrome configurations cause latency spikes.
The batch $p_{99}$ deadline is met only at $d \leq 5$ for a single CPU core, compared to $d \leq 7$ for the mean.

\paragraph*{Claim~5: Surface codes and qLDPC codes impose different system requirements.}
High-rate codes such as qLDPC codes can reduce data-qubit overhead, but introduce non-local connectivity requirements, additional ancilla and measurement resources, deeper syndrome-extraction circuits, and less mature real-time decoder pipelines.
This confirms that qLDPC codes require fundamentally different decoder architectures and hardware backends from surface-code systems.

\paragraph*{Claim~6: Architecture and systems engineering will be the next differentiators.}
We proposed a six-layer reference architecture (Chapter~\ref{sec:architecture}) with heterogeneous compute dispatch, speculative decoding, and backlog management.
The architecture also generalizes the decoder input from a bare binary syndrome stream to a structured readout interface that can carry confidence, erasure/loss, timing, and hardware-layout metadata.
These capabilities require computer systems expertise instead of quantum physics breakthroughs.
The scaling analysis (Section~\ref{sec:scaling}) shows that hardware acceleration is the primary scaling lever.

\subsection*{Limitations and future work}

Our benchmarks use a uniform depolarizing noise model, which captures the essential error-threshold behavior but omits effects present in real devices: coherent errors, leakage, mid-circuit readout crosstalk, and temporal fluctuations in error rates from two-level system noise~\cite{google2025threshold}.
Real-device noise typically produces correlated error patterns that generate heavier decoding loads than depolarizing predictions, particularly in the tail of the decoding time distribution that governs hard-deadline compliance.
The six-layer reference architecture is a design proposal, but end-to-end validation requires co-designed QPU, control electronics, and decoder stack working together, where different hardware platforms will require different stack implementations.

Key open problems that extend beyond the scope of this paper include:
\begin{itemize}
  \item \textbf{Hardware-accelerated qLDPC decoding.} BP+OSD is a widely used decoder for qLDPC codes, but faster specialized decoders, BP variants, and hardware-oriented implementations are active research directions.
  Meanwhile, GPU-accelerated BP+OSD implementations have not yet been fully explored.
  \item \textbf{Standardized decoder interfaces.} No community standard exists for syndrome packet formats or decoder APIs, limiting portability and benchmarking comparability. 
  This interface will differ across hardware platforms. 
  \item \textbf{Platform-specific real-time benchmarks.} The benchmarks in this paper are primarily decoder-centric.
  Future benchmarks should include trapped-ion and neutral-atom closed-loop latency budgets, including readout classification, metadata transport, Pauli-frame update, feedforward, QCCD shuttling or atom movement, and erasure-aware decoding.
  \item \textbf{End-to-end implementation at scale.} The synchronization regime analysis and dynamic dispatch policies that adjust decoder selection based on real-time syndrome density or device drift remain uncharacterized experimentally.
\end{itemize}

\bibliography{ref}

\end{document}